\documentclass[pra,aps,twocolumn,eqsecnum,showpacs]{revtex4}

\usepackage{dcolumn}
\usepackage{amsmath}
\usepackage{graphics}

\begin{document}

\title{Multimode entanglement of light and atomic ensembles via
off-resonant coherent forward scattering}
\author{D.V. Kupriyanov, O.S. Mishina, and I.M. Sokolov }
\affiliation{Department of Theoretical Physics, State Polytechnic
University, 195251, St.-Petersburg, Russia}%
\email{Kupr@quark.stu.neva.ru}%
\author{B. Julsgaard, E.S.Polzik}
\affiliation{QUANTOP - Danish Quantum Optics Center, Niels Bohr
Institute, 2100 Copenhagen, Denmark}%
\email{polzik@nbi.dk}%

\date{\today }

\begin{abstract}
Quantum theoretical treatment of coherent forward scattering of
light in a polarized atomic ensemble with an arbitrary angular
momentum is developed. We consider coherent forward scattering of
a weak radiation field interacting with a realistic multi-level
atomic transition. Based on the concept of an effective
Hamiltonian and on the Heisenberg formalism, we discuss the
coupled dynamics of the quantum fluctuations of the polarization
Stokes components of propagating light and of the collective spin
fluctuations of the scattering atoms. We show that in the process
of coherent forward scattering this dynamics can be described in
terms of a polariton-type spin wave created in the atomic sample.
Our work presents a general example of entangling process in the
system of collective quantum states of light and atomic angular
momenta, previously considered only for the case of spin
$\small{1/2}$ atoms. We use the developed general formalism to
test the applicability of spin $\small{1/2}$ approximation for
modelling the quantum non-demolishing measurement of atoms with a
higher angular momentum.
\end{abstract}

\pacs{03.67.Mn, 34.50.Rk, 34.80.Qb, 42.50.Ct}%
\maketitle%


\section{Introduction}
Various optical phenomena associated with the optical pumping
process, which have been comprehensively studied since the 60's
and described in many aspects in a famous review \cite{Hpr} by
W.Happer, is revived nowadays in a new form within the field of
quantum information and quantum computing. The paramagnetic ground
states of macroscopic atomic spin subsystems are considered now
convenient physical objects for mapping and storing the quantum
information in the quantum states of their collective angular
momenta. The Faraday-type interference scheme was proposed for
spin squeezing \cite{K1} and for quantum communication between
atomic ensembles \cite{DCZP}. The proposed ideas were realized in
spin squeezing \cite{K2} and in the entanglement \cite{JKP}
experiments, where a quantum measurement on the light
forward-scattered from atomic ensembles was used. The same kind of
off-resonant forward scattering combined with a quantum feedback
was used in the recent demonstration of quantum memory for light
\cite{H}. Applications to various quantum information protocols
including a cat state generation \cite{J} and quantum cloning of
light onto atoms \cite{cloning} have been proposed. Efficient
generation of entanglement via multi-pass interaction have been
also proposed \cite{Polzik}. Theoretical modelling in the above
mentioned papers was concerned with collective canonical variables
for atoms which can be conveniently introduced for spin
$\small{1/2}$ atoms. However, the actual experiments were
conducted using states with higher angular momenta. Hence a
theoretical model describing off-resonant interaction of light
with realistic atoms with the angular momentum higher then
$\small{1/2}$ is required.

It is worth noting that the effect of a random Faraday rotation
due to atomic fluctuations was discussed and observed first in
Refs.\cite{AlZp,GlPl} more than twenty years ago in the context of
demonstrating the advantages of light-beating method in atomic
spectroscopy. The importance of atomic polarization was later
discussed in Refs.\cite{KpSk,BKS} where it was shown that the
collective spin polarization of an atomic ensemble could
essentially modify the quantum statistics of the outgoing
off-resonant probe radiation. Quantum statistics of atomic spin
variables was first experimentally observed via off-resonant
forward scattering in \cite{Hald}.

In the present paper we develop the quantum theory of coherent
forward scattering on an ensemble of polarized atoms with an
arbitrary angular momentum. We discuss the physical conditions
under which the forward scattering can be properly described in
terms of the effective Hamiltonian, generalizing the semiclassical
approach \cite{Hpr}, so that both the field and atomic subsystems
are treated as quantum objects. Based on the effective Hamiltonian
we derive the Heisenberg-type equations of motion written for the
spatial distribution of the Stokes variables of the field and of
the collective angular momentum of atoms. The solution shows how
the entanglement of the quantum fluctuations of light and atomic
subsystems is formed. We show that the entangling process can be
understood in terms of a polariton-type spin wave induced in the
atomic sample. We support our discussion by numerical simulations,
assuming the conditions close to those of the recent experiments
\cite{JKP}.

The paper is organized as follows. In section \ref{s2} we review
the process of coherent forward scattering to show how the field
Heisenberg operators are transformed in interaction with
multi-atom ensemble consisting of atoms with an arbitrary Zeeman
structure and for an off-resonant excitation on a dipole allowed
optical transition. In section \ref{s3} we extend the Heisenberg
formalism to the atomic subsystem and introduce the effective
Hamiltonian responsible for the coupled dynamics of an infinite
number of variables associated with the local polarization of the
weak probe light and with individual atomic spins. The effective
Hamiltonian is further transformed in section \ref{s4} into
another form by introducing the more convenient for the
polarization sensitive processes irreducible tensor formalism. In
section \ref{s5} we perform mesoscopic averaging to arrive at the
wave-type equations for the collective quantum variables of the
light and atomic subsystems. These equations are solved and
discussed in section \ref{s6} in the context of the quantum
entangling process. In particular they are used to test the
applicability of the spin one-half approximation for modelling the
quantum non-demolishing measurement in an ensemble of alkali atoms
with a higher angular momentum.

\section{Coherent forward scattering of light in the Heisenberg
picture}\label{s2}

Consider a system of $N$ identical atoms located in a finite
volume and scattering coherent light of frequency $\omega $. The
wave of light incident on the atomic ensemble is assumed to be
weak enough that possible saturation effects in its interaction
with the atoms are negligible. The atomic ensemble, in general,
represents an optically long (thick in refraction, but thin in
absorption) medium for multiple off resonance scattering, but, on
the average, the atoms are separated by distances much larger than
the wavelength $\lambdabar$. Thus each atom is located, on
average, in the radiation zone of its neighbors. The interaction
of the atoms with incident and multiply-scattered light is assumed
to be of the dipole type and a proper description of multiple
scattering has to be restricted to the rotating wave
approximation.

To follow as clearly as possible the analogy between classical and
quantum descriptions of the coherent forward scattering it seems
convenient to solve the quantum problem in the Heisenberg picture.
In such an approach the basic process is the transformation of the
electric field operators in a single scattering event. This
process is reviewed in appendix \ref{s2.1}, as an example of the
transformation of the operator of the free electric field of a
moving atom. As shown there, the basic result can be written in
terms of the Heisenberg-type microscopic Maxwell equation, where
polarization response-operator is given by a single point-like
scatterer. This allows us to make a subsequent generalization to
the situation of an arbitrary number of atoms scattering light
coherently in the forward direction.

We begin our discussion from the single particle microscopic
problem, described by equation (\ref{2.10})(see the appendix
\ref{s2.1} for details). For a system of $N$ identical scatterers
randomly located in space it would be not so easy to generalize
(\ref{2.10}) to a multi-particle form allowing to follow precisely
the Heisenberg dynamics. For simplicity we restrict the discussion
to the case of the plain incident wave, as in the experiments with
gas cells \cite{JKP,H}. For a transparent medium consisting of $N$
scatterers light is scattered in any direction other than the
forward direction only incoherently. This means that for optically
thin and transparent medium the probability to observe a photon
randomly scattered in any solid angle different from the direction
of light propagation is just a sum of partial probabilities of
independent scattering by each atom. In the Heisenberg formalism
this means that the non-forward propagating operator waves created
in the medium by single or multiple scattering events are
superposed with random phases to a roughly zero sum. The outgoing
flux associated with incoherent scattering is proportional to the
total number of atoms, but stays small because of a negligible
value of the off-resonant cross section. At the same time for
light propagating in the forward direction in transparent medium
there is a very strong coherent enhancement of the scattering
process. Since the Doppler shift in the forward scattered modes
disappears, see (\ref{2.8}), and the Raman shift, caused by Zeeman
splitting, does not noticeably change the phase of these modes,
the transmitted light reveals strong coherent superposition of all
partial contributions associated with single and multiple
scattering. For discussion of the off-resonant light-atom coherent
scattering beyond the plain wave approximation see \cite{KK,MP}.

Let us subdivide the joint dynamics of light and atomic subsystems
into time increments, during which the number of accumulated
incoherent scattering events is much less than the total number of
atoms. This simplifies our analysis and allows to consider only
the modes coherently scattered in the forward direction. In the
examples we are going to discuss below this type of scattering is
additionally stimulated by the propagating mode, which has the
quasi-classical nature. Within the rotating wave approximation we
introduce a carrier frequency $\bar{\omega}$, coinciding with the
average frequency of propagating light, and define an averaged
wavenumber $\bar{k}=\bar{\omega}/c$ associated with the full set
of the modes propagating in the $z$-direction. Then the spectral
bandwidth $\Delta\omega$ of the continuum of the field modes,
contributing to the quantum operator expansion, should be chosen
to be less than $\bar{\omega}$. The actual spectrum of the
incident radiation, centered at $\bar{\omega}$, has a narrower
bandwidth $\delta\omega$, which is much less than the detuning
from the resonance $\Delta=\bar{\omega}-\omega_0$, where
$\omega_0$ is the resonance frequency of the non-perturbed optical
transition. We will also assume that $|\Delta|$ is much greater
than natural linewidth $\gamma_n$, Doppler shift $\bar{k}\bar{v}$,
and Raman shift $\omega_{m'm}$. We neglect the imaginary shift
associated with the natural linewidth in the energy denominators.
The dissipation process associated with the spontaneous decay is
neglected in our discussion; the assumptions justifying this are
discussed in section \ref{s6.c}. For an analysis of the role of
spontaneous emission for off-resonant entangling interaction see
\cite{Polzik}.

Under the above mentioned assumpltions the basic expression for
the polarization of any $a$-th atom contributing to the forward
scattering (\ref{2.15}) can be rewritten as follows
\begin{equation}
\hat{P}^{(a,+)}_{i}(z,t)\;=\;\frac{1}{S_0}\;%
\hat{\alpha}^{(a)}_{ij}(\bar{\omega},t)\,\delta(z-z_a(t))\,%
\hat{E}^{(+)}_j(z,t)%
\label{2.16}%
\end{equation}
where
\begin{equation}
\hat{\alpha}^{(a)}_{ij}(\bar{\omega},t)\;=\;%
\sum\limits_{m,m^{\prime}}\sum\limits_{n}%
\frac{(d^a_i)_{m^{\prime }n}\,%
(d^a_j)_{nm}}%
{-\hbar(\bar{\omega} -\omega _{nm}-\bar{k}{v}_{za})}%
|m'\rangle\langle m|^{(a)}(t)%
\label{2.17}%
\end{equation}%
is the polarizability tensor and $S_0$ is the crossection area of
the light beam propagating through the medium in $z$-direction.
Other notations are specified in appendix \ref{s2.1}. Expression
(\ref{2.17}) gives the instantaneous value of the polarizability
tensor for $a$-th atom, which depends on its exact Heisenberg
evolution to the moment $t$. The Heisenberg operator of the
electric field on the right hand side of equation (\ref{2.16}) is
assumed to be dressed in the perturbations by other atoms located
in front of any selected $a$-th atom but non-perturbed by the
selected atom itself. Here we use Cartesian coordinates in the
tensor notation and sum over each repeated index. The indices
$i,j$ take values $x$ or $y$.

Let us consider a thin mesoscopic layer located between $z$ and
$z+\Delta z$ planes and containing a large number of atoms. Such
an atomic sub-ensemble will  scatter incoming field coherently in
the forward direction by the collective polarization
\begin{equation}
\hat{\mathbf{P}}^{(+)}(z,t)\;=\;%
\sum_{a=1}^{N}\,\hat{\mathbf{P}}^{(a,+)}(z,t)
\label{2.18}%
\end{equation}%
where $z$ coordinate is confined inside the layer $(z,z+\Delta
z)$. If the medium is split in a number of thin layers along the
propagation direction the light beam would be subsequently
scattered by this layers in the forward direction. Then the
"coarse-grained" dynamics of the field operators can be described
by the following Heisenberg-type macroscopic Maxwell equation
\begin{equation}
\frac{\partial^2}{\partial z^2}\hat{\mathbf{E}}^{(+)}(z,t)\,-\,%
\frac{1}{c^2}\frac{\partial^2}{\partial t^2}%
\hat{\mathbf{E}}^{(+)}(z,t)\;=\;%
\frac{4\pi}{c^2}\frac{\partial^2}{\partial t^2}%
\hat{\mathbf{P}}^{(+)}(z,t)%
\label{2.19}%
\end{equation}%
where the polarization operator $\hat{\mathbf{P}}^{(+)}(z,t)$ is
subsequently given by expressions (\ref{2.16}) - (\ref{2.18}).

We conclude this section by the following remark. The derived
macroscopic Maxwell equation is coupled to the corresponding
Heisenberg equations governing the dynamics of the atomic
subsystem, see below, and it cannot be extended up to an arbitrary
time $t$. Its validity is restricted by ignoring the dissipation
process of incoherent scattering. That is why the averaging of the
operator polarizability (\ref{2.17}) gives only the refraction
part of the real polarizability tensor of normal Maxwell
equations. We will further discuss the self-consistency of the
dynamical approach in section \ref{s6.c}.

\section{Dynamics of atomic subsystem}\label{s3}

In this section we discuss how the atomic variables are modified
via the interaction with the forward propagating light. We
describe the dynamics of slowly varying ground state operators,
considering at first a single atom illuminated with an arbitrary
quantized optical field. Then we generalize the problem to a
macroscopic ensemble and introduce atomic collective variables.

\subsection{Dynamics of a single atom coupled to
off-resonant field}

Consider an atom at the origin of the coordinate frame at the
initial moment of time and drifting in space in such a way that
during the scattering event it moves much less than a wavelength.
Let us define an arbitrary dyadic-type operator for the ground
state of this atom
\begin{equation}
\hat{T}\;=\;|m'\rangle\langle m|\;\equiv\;%
|j_0m'\rangle\langle j_0m|%
\label{3.1}%
\end{equation}%
where by the second equality we specify the atomic state more
precisely and introduce the atomic quantum numbers: $j_0$ and
$m,m'$ are the ground state angular momentum and its projections
on the direction of an external magnetic field oriented along the
$Z$ axis, which in general is different from the direction of the
light propagation. Based on a perturbation theory we can expand
the corresponding Heisenberg operator up to the second order as
follows
\begin{equation}
\hat{T}(t)\;=\;\hat{T}_0(t)\,+\,\hat{T}_2(t)\,+\,\ldots%
\label{3.2}%
\end{equation}%
where
\begin{equation}
\hat{T}_0(t)\;=\;%
e^{i\omega_{m'm}t}\,|m^{\prime }\rangle \langle m|%
\label{3.3}%
\end{equation}%
is the operator in the interaction representation and the second
term
\begin{eqnarray}
\hat{T}_{2}(t)&=&-\frac{1}{\hbar^{2}}\,%
\int_{0}^{t}dt^{\prime \prime}%
\int_{t^{\prime \prime }}^{t}dt^{\prime }\,%
\nonumber\\%
&&\times\left[ \hat{\mathbf{d}}(t^{\prime \prime })\,%
\hat{\mathbf{E}}_{0}(t^{\prime \prime }),\,%
\left[ \hat{\mathbf{d}}(t^{\prime })\,%
\hat{\mathbf{E}}_{0}(t^{\prime }),\,%
\hat{T}_{0}(t)\right] \right]%
\label{3.4}%
\end{eqnarray}%
is the second order correction induced by a dipole-type optical
interaction, see (\ref{2.2}).

The integral (\ref{3.4}) should be evaluated in the rotating wave
approximation by keeping only the leading terms in the limit $t\gg
|\Delta|^{-1}$. These terms can only depend on normally ordered
products of the creation and annihilation field operators. In
general such products expand over all the spatial modes, but in
reality only those modes, which will not vanish after the
averaging over the initial state, are important. These are planar
modes propagating along the $z$-axis. Taking into account only the
propagating modes, the expansion (\ref{3.2}) can be rewritten as
follows
\begin{equation}
\hat{T}(t)\;\approx\;\hat{T}_0(t)\,+\,\int_0^t dt'\;%
\frac{i}{\hbar}\left[\hat{\cal{H}}_{eff}(t'),\,%
\hat{T}_0(t)\right]\,+\,\ldots%
\label{3.5}%
\end{equation}%
where we introduce an effective interaction Hamiltonian in the
interaction representation
\begin{equation}
\hat{\cal{H}}_{eff}(t)\;=\;%
-\hat{E}^{(-)}_{i}(z,t)\,%
\hat{\alpha}_{ij}(\bar{\omega},t)\,%
\hat{E}^{(+)}_{j}(z,t)\,%
\label{3.6}%
\end{equation}
Strictly speaking the electric field operators and the operator of
the atomic polarizability tensor should be understood here as
defined in the interaction representation and marked by the index
zero. For the electric field these operators
$\hat{\mathbf{E}}^{(\pm)}(z,t)$ are given by the expansion
(\ref{2.1}), defined at $z\to 0$, where we keep only the forward
propagating modes. The operator of the polarizability tensor
$\hat{\alpha}_{ij}(\bar{\omega},t)$ is given by Eq.(\ref{2.17}),
transformed to the interaction representation, with $\bar{\omega}$
and $\bar{k}=\bar{\omega}/c$ being the carrier frequency and
wavenumber of the modes interacting with an atom. As in the
Eq.(\ref{2.16}) the tensor indices inside the definition of the
effective Hamiltonian (\ref{3.6}) can run only two projections
either $x$ or $y$.

Note that in our derivation of the effective Hamiltonian we used a
rather short time increment $t$ consistent with the perturbation
theory approach; with the assumption that the atom does not
noticeably change its location during the scattering event. Then
there would be no difference between the interaction and the
Heisenberg representations in the evaluation of the integral
(\ref{3.4}) and in introducing the effective Hamiltonian
(\ref{3.6}). Based on a general principle of dynamical evolution
we can straightforwardly generalize Eq.(\ref{3.6}) up to an
arbitrary moment in time if we substitute there all the operators
in the Heisenberg representation. We should also take into account
a classical drift of the atom in space and consider $z=z(t)$
coordinate as its actual location at moment $t$.

\subsection{Generalization to a multi-atom ensemble}

Consider now an ensemble consisting of many atoms scattering
incident light coherently in the forward direction. Although the
most interesting polariton-type solutions obtained further in the
paper are destroyed by atomic motion, we include the motion in our
model for the sake of generality. However we ignore any possible
correlations between their spatial motion and the internal state
evolution. Then each atom is in the environment of the field
scattered by the atoms located in front of it and coupled with
such a dressed field via the partial effective Hamiltonian
(\ref{3.6}). The full effective Hamiltonian for the whole ensemble
interacting with the propagating field in the Heisenberg
representation is given by
\begin{equation}
\hat{\cal{H}}_{eff}(t)\;=\;%
-\sum_{a=1}^N\;%
\hat{E}^{(-)}_{i}(z_a(t),t)\;%
\hat{\alpha}_{ij}^{(a)}(\bar{\omega},t)\;%
\hat{E}^{(+)}_{j}(z_a(t),t)\,%
\label{3.7}%
\end{equation}%
We preserve here the notation for the full effective Hamiltonian,
which was used in the previous equation in the case of one atom.

Then operator (\ref{3.1}) considered for each atom of the ensemble
satisfies the following Heisenberg equation
\begin{equation}
\dot{\hat{T}}{}^{(a)}(t)\;=\;%
i\omega_{m'm}\,\hat{T}^{(a)}(t)\;+\;%
\frac{i}{\hbar}\left[\hat{\cal{H}}_{eff}(t),\,\hat{T}^{(a)}(t)\right]%
\label{3.8}%
\end{equation}
where $a=1\div N$.

If we take into account the commutation relation between
Heiesenberg operators of the electric field propagating in the
forward direction, which are given by
\begin{equation}
\left[\hat{E}^{(-)}_{k}(z',t),\,\hat{E}^{(+)}_{i}(z,t)\right]\;=\;%
-\frac{2\pi\hbar\bar{\omega}}{S_0}\,\delta_{ki}\,\delta(z'-z)%
\label{3.9}%
\end{equation}%
then the Maxwell equation (\ref{2.19}) can be also rewritten in
terms of the effective Hamiltonian
\begin{widetext}%
\begin{equation}
\frac{\partial^2}{\partial z^2}\hat{E}_i^{(+)}(z,t)\,-\,%
\frac{1}{c^2}\frac{\partial^2}{\partial t^2}%
\hat{E}_i^{(+)}(z,t)\;=\;%
-\frac{2\bar{\omega}}{\hbar c^2}\,%
\left[\hat{\cal{H}}_{eff}(t),\,\hat{E}_i^{(+)}(z,t)\right]%
\label{3.10}%
\end{equation}%
\end{widetext}%
This operator equation shows that, in turn, the state of the field
existing at each spatial point of the medium is defined by the
atomic operators in the Heisenberg representation, according to
their spatial distribution in space at a given time $t$.

Let us make the following remark concerning the definition of
$\delta$-functions in (\ref{3.9}) and in (\ref{2.16}) and the
validity of the dynamical equations in the form (\ref{3.10}). Our
description of the quasi-resonant radiation propagating in
disordered medium in the forward direction is based on the
rotating wave approximation. According to this approximation the
spectral bandwidth of the field $\Delta\omega$ is assumed to be at
least less than carrier frequency $\bar{\omega}$. Such a
truncation of the infinite field continuum makes possible to
consider the commutation relations of truly Heisenberg operators
in the form (\ref{3.9}). Thus the $\delta$-functions here and in
(\ref{2.16}) should be correctly understood as distributed in a
small mesoscopic area of the order $c/\Delta\omega$. This spatial
scale is obviously longer than $\lambdabar$ but still much shorter
than the sample size or than any internal macroscopic scale
associated with macroscopic susceptibilities of the medium. It is
also important to think about equations (\ref{3.10}) as on
approximation for more general multi-mode Heisenberg-Langevin type
equation, where the damping processes associated with incoherent
scattering would be taken into consideration. As was mentioned
before, the validity of the output Maxwell equations (\ref{2.19}),
(\ref{3.10}) is actually restricted by the narrow spectral domain
$\delta\omega$ associated with the spectrum of the probe light.
But in the process of transposing the field operators by means of
the commutation rule (\ref{3.9}) the whole field spectrum
$\Delta\omega$ should be taken into account.

The coupled equations (\ref{3.8}) and (\ref{3.10}) considered
together is the main result of this section. They reveal the joint
dynamics of the field and atomic subsystems interacting in the
limit of non-saturating off-resonant optical excitation. Being an
example of Heisenberg equations they are valid for arbitrary
initial quantum state of the field and of the atomic ensemble. The
main restriction comes from the model of lossless coherent
scattering. But even with such a simplification these equations
are quite complicated since they are operator equations for an
{\it infinite number} of the field and atomic variables. In the
following sections we identify and discuss special conditions when
equations (\ref{3.8}) and (\ref{3.10}) could be converted to a
{\it finite number} of Heisenberg-type equations for collective
variables of the field and atomic subsystems.

\section{Representation of irreducible components. Stokes
operators of the electromagnetic field}\label{s4}

\subsection{Transformation of the effective Hamiltonian to the
irreducible representation}

The set of dyadic-type operators (\ref{3.1}) for each atom of the
ensemble can be replaced with another set of operators
\begin{eqnarray}
\hat{T}_{KQ}^{(a)}&=&\sqrt{\frac{2K+1}{2j_0+1}}\,%
\sum_{m',m}\,C_{j_0m\,KQ}^{j_0m'}\,|j_0m'\rangle\langle j_0m|^{(a)}%
\nonumber \\%
|j_0m'\rangle\langle j_0m|^{(a)}&=&\sum_{KQ}\,%
\sqrt{\frac{2K+1}{2j_0+1}}\,%
C_{j_0m\,KQ}^{j_0m'}\,\hat{T}_{KQ}^{(a)}%
\label{4.1}%
\end{eqnarray}%
Being the linear combination  of original operators weighted with
Clebsh-Gordan coefficients $C_{\ldots\,\ldots}^{\ldots}$, the
projectors $\hat{T}_{KQ}^{(a)}$ become irreducible tensor
operators, which possess the simplest properties with respect to
rotational transformations, see \cite{VMK}. The linear
transformations (\ref{4.1}) are consistent with any coordinate
frame. Let us abandon the choice of the frame with $Z$-axis along
magnetic field, used in the basic definition (\ref{3.1}), and
return to the original frame with the $z$-axis along the
propagating beam, which is more natural for the further discussion
of the effective Hamiltonian approach.

For large frequency detuning the Zeeman splitting in the energy
denominators can be neglected compared with the average detuning
$\Delta=\bar{\omega}-\omega_{jj_0}$, where
$\omega_{jj_0}=\omega_0$ is the transition frequency between the
ground and excited states characterized by the angular momenta
$j_0$ and $j$ respectively. Then the effective Hamiltonian can be
expressed as the sum of three terms
\begin{equation}
\hat{\cal{H}}_{eff}(t)\;=\;%
\hat{\cal{H}}_{eff}^{(0)}(t)\,+\,\hat{\cal{H}}_{eff}^{(1)}(t)\,%
+\,\hat{\cal{H}}_{eff}^{(2)}(t)%
\label{4.2}%
\end{equation}%
where dependence on time emphasizes that the contributing atom and
field operators are taken in the Heisenberg representation.

The first term in (\ref{4.2}) couples the atomic population of the
whole Zeeman multiplet and their longitudinal alignment with the
full photon flux of propagating light
\begin{eqnarray}
\hat{\cal{H}}_{eff}^{(0)}(t)&=&%
-\frac{2\pi\hbar\bar{\omega}}{S_0\,c}\,\left[\alpha_0(\bar{\omega})\,%
\sum_{a=1}^{N}\,\hat{T}_{00}^{(a)}(t)\right.%
\phantom{\frac{1}{\sqrt{6}}\,\alpha_2(\bar{\omega})\,\sum_{a=1}^{N}\,}%
\nonumber\\%
&&\left.+\,\frac{1}{\sqrt{6}}\,\alpha_2(\bar{\omega})\,\sum_{a=1}^{N}\,%
\hat{T}_{20}^{(a)}(t)\right]\;%
\hat{\Xi}_0\left(z_a(t),t\right)%
\label{4.3}%
\end{eqnarray}%
where
\begin{equation}
\hat{\Xi}_0(z,t)\;=\;%
\frac{S_0\,c}{2\pi\hbar\,\bar{\omega}}\,%
\hat{\mathbf{E}}^{(-)}(z,t)\,\hat{\mathbf{E}}^{(+)}(z,t)%
\label{4.4}%
\end{equation}%
is the Stokes operator of the total photon flux at the spatial
point $z$. The isotropic polarizability of an atom is given by
\begin{equation}
\alpha_0(\bar{\omega})\;=\;%
\frac{1}{3\sqrt{2j_0+1}}\;%
\frac{|d_{j_0j}|^2}{-\hbar(\bar{\omega}-\omega_{jj_0})}%
\label{4.5}%
\end{equation}%
where $d_{j_0j}$ is the reduced matrix element of the atomic
dipole moment for $j_0\to j$ transition. The alignment component
of the atomic polarizability is given by Eq.(\ref{4.13}) below.

The second term in Eq.(\ref{4.2}) couples the gyrotropic or
orientation component of the atomic ensemble with the Stokes
component responsible for circular polarization of propagating
light
\begin{equation}
\hat{\cal{H}}_{eff}^{(1)}(t)\;=\;%
\frac{2\pi\hbar\bar{\omega}}{S_0\,c}\,\alpha_1(\bar{\omega})\,%
\sum_{a=1}^{N}\,\hat{T}_{10}^{(a)}(t)\;\hat{\Xi}_2\left(z_a(t),t\right)%
\label{4.6}%
\end{equation}%
where
\begin{eqnarray}
\hat{\Xi}_2(z,t)&=&%
\frac{S_0\,c}{2\pi\hbar\,\bar{\omega}}\,%
\left[\hat{E}_R^{(-)}(z,t)\,\hat{E}_R^{(+)}(z,t)\right.%
\nonumber\\%
&&\phantom{\frac{S_0\,c}{2\pi\hbar\,\bar{\omega}}\,}%
\left.-\,\hat{E}_L^{(-)}(z,t)\,\hat{E}_L^{(+)}(z,t)\right]%
\label{4.7}%
\end{eqnarray}%
is the Stokes operator associated with circular polarization. It
is defined in terms of the photon flux operators at a spatial
point $z$ and shows imbalance between the right-hand ($R$) and the
left-hand ($L$) polarizations of the field. The orientational
polarizability of an atom is given by
\begin{equation}
\alpha_1(\bar{\omega})\;=\;%
(-)^{j+j_0}\frac{1}{\sqrt{2}}\,%
\left\{\begin{array}{ccc}1&1&1\\j_0&j_0&j\end{array}\right\}\,%
\frac{|d_{j_0j}|^2}{-\hbar(\bar{\omega}-\omega_{jj_0})}%
\label{4.8}%
\end{equation}%
As follows from the Maxwell equation (\ref{3.10}) in the classical
limit, the term $\hat{\cal{H}}_{eff}^{(1)}$ completely defines the
Faraday rotation or other gyrotropy effects existing in a bulk
medium.

The third term in Eq.(\ref{4.2}) couples the alignment components
of the atomic ensemble with the remaining two linear polarized
type Stokes components of propagating light
\begin{eqnarray}
\hat{\cal{H}}_{eff}^{(2)}(t)&=&%
\frac{2\pi\hbar\bar{\omega}}{S_0\,c}\,\alpha_2(\bar{\omega})\,%
\sum_{a=1}^{N}\,%
\left[\hat{T}_{xy}^{(a)}(t)\;\hat{\Xi}_3\left(z_a(t),t\right)\right.%
\nonumber\\%
&&\left.+\,\hat{T}_{\xi\eta}^{(a)}(t)\;\hat{\Xi}_1\left(z_a(t),t\right)\right]%
\label{4.9}%
\end{eqnarray}
where
\begin{eqnarray}
\hat{T}_{xy}^{(a)}(t)&=&\frac{1}{2}\left[\hat{T}_{2-2}^{(a)}(t)\,+\,%
\hat{T}_{22}^{(a)}(t)\right]%
\nonumber\\%
\hat{T}_{\xi\eta}^{(a)}(t)&=&%
-\frac{1}{2i}\left[\hat{T}_{2-2}^{(a)}(t)\,-\,%
\hat{T}_{22}^{(a)}(t)\right]%
\label{4.10}%
\end{eqnarray}%
The Stokes operator $\hat{\Xi}_3(z,t)$, showing imbalance between
the photon fluxes of the modes linearly polarized along $x$ and
$y$ axes, is given by
\begin{eqnarray}
\hat{\Xi}_3(z,t)&=&%
\frac{S_0\,c}{2\pi\hbar\,\bar{\omega}}\,%
\left[\hat{E}_x^{(-)}(z,t)\,\hat{E}_x^{(+)}(z,t)\right.%
\nonumber\\%
&&\phantom{\frac{S_0\,c}{2\pi\hbar\,\bar{\omega}}\,}%
\left.-\,\hat{E}_y^{(-)}(z,t)\,\hat{E}_y^{(+)}(z,t)\right]%
\label{4.11}%
\end{eqnarray}%
The Stokes operator $\hat{\Xi}_1(z,t)$, showing imbalance between
photon fluxes of the modes linearly polarized along $\xi$ and
$\eta$ axis rotated with respect to $x$ and $y$ directions by
$\pi/4$ angle, is given by
\begin{eqnarray}
\hat{\Xi}_1(z,t)&=&%
\frac{S_0\,c}{2\pi\hbar\,\bar{\omega}}\,%
\left[\hat{E}_\xi^{(-)}(z,t)\,\hat{E}_\xi^{(+)}(z,t)\right.%
\nonumber\\%
&&\phantom{\frac{S_0\,c}{2\pi\hbar\,\bar{\omega}}\,}%
\left.-\,\hat{E}_\eta^{(-)}(z,t)\,\hat{E}_\eta^{(+)}(z,t)\right]%
\label{4.12}%
\end{eqnarray}
The alignment component of the atomic polarizability is defined as
follows
\begin{equation}
\alpha_2(\bar{\omega})\;=\;(-)^{j+j_0+1}\,%
\left\{\begin{array}{ccc}1&1&2\\j_0&j_0&j\end{array}\right\}\,%
\frac{|d_{j_0j}|^2}{-\hbar(\bar{\omega}-\omega_{jj_0})}%
\label{4.13}%
\end{equation}%
As follows from the Maxwell equation (\ref{3.10}) considered in
its classical limit the term $\hat{\cal{H}}_{eff}^{(2)}$ is
responsible for the optical birefringence effects with respect to
either $x$ and $y$ or $\xi$ and $\eta$ directions.

\subsection{Dynamical equations driven by the effective Hamiltonian}

Equation (\ref{3.8}) in the irreducible representation transforms
into
\begin{eqnarray}
\dot{\hat{T}}{}_{KQ}^{(a)}(t)&=&%
\frac{i}{\hbar}\left[\hat{H}_0(t),\,\hat{T}_{KQ}^{(a)}(t)\right]%
\nonumber\\%
&&+\;\frac{i}{\hbar}\left[\hat{\cal{H}}_{eff}(t),\,%
\hat{T}_{KQ}^{(a)}(t)\right]%
\label{4.14}%
\end{eqnarray}
where $\hat{H}_0$ is the unperturbed Hamiltonian responsible for
the interaction with the external magnetic field. Recall here that
in general the direction of the magnetic field  does not coincide
with the $z$-axis. The set of equations (\ref{4.14}) for $a=1\div
N$ is equivalent to the set of equations (\ref{3.8}) but it is
written for more convenient physical observables. The irreducible
components of a low rank, which contribute to the effective
Hamiltonian and couple to the Stokes observables of the
propagating light, allow for a more transparent interpretation
than the original projector operators (\ref{3.1}). The zero rank
component $\hat{T}_{00}^{(a)}(t)$ is the Heisenberg operator of
the total population of an a-th atom in its ground state. It can
be straightforwardly verified that the right hand size of
(\ref{4.14}) is equal to zero in this case and the zero rank
irreducible operator is, in fact, the identity operator. The first
rank component $\hat{T}_{1Q}^{(a)}(t)$, being the Heisenberg
operator of atomic orientation, is equivalent to the vector of the
atomic angular momentum. This vector undergoes dynamical evolution
(regular precession and coupling with the field variables) caused
by the perturbation of the atomic ground state by the propagating
light. The second rank component $\hat{T}_{2Q}^{(a)}(t)$, being
the Heisenberg operator of atomic alignment, is equivalent to the
ground state quadrupole moment of the atom. The quadropole tensor
also undergoes the dynamical evolution caused by interaction with
the propagating light. Other higher rank irreducible components
are important only as long as their evolution is dynamically
coupled with the evolution of the lower rank components in the
complete set of Eqs.({\ref{4.14}}).

The equations (\ref{4.14}) have to be solved together with the
Maxwell equations. Let us introduce the slowly varying amplitudes
of the Heisenberg field operators
\begin{eqnarray}
\hat{E}_i^{(+)}(z,t)&=&\hat{\epsilon}_i(z,t)\,%
e^{-i\bar{\omega}t\,+\,i\bar{k}z}%
\nonumber\\%
\hat{E}_i^{(-)}(z,t)&=&\hat{\epsilon}^\dagger_i(z,t)\,%
e^{i\bar{\omega}t\,-\,i\bar{k}z}%
\label{4.15}%
\end{eqnarray}
By substituting these expressions into equation (3.10) and into
its Hermitian conjugated form, we obtain the following first order
differential equations for slowly varying amplitudes
\begin{eqnarray}
\left[\frac{\partial}{\partial z}+\frac{1}{c}%
\frac{\partial}{\partial t}\right]%
\hat{\epsilon}_i(z,t)&=&%
\frac{i}{\hbar c}\left[\hat{\cal{H}}_{eff}(t),\,%
\hat{\epsilon}_i(z,t)\right]%
\nonumber\\%
\left[\frac{\partial}{\partial z}+\frac{1}{c}%
\frac{\partial}{\partial t}\right]%
\hat{\epsilon}^\dagger_i(z,t)&=&%
\frac{i}{\hbar c}\left[\hat{\cal{H}}_{eff}(t),\,%
\hat{\epsilon}^\dagger_i(z,t)\right]%
\label{4.16}%
\end{eqnarray}%
In such Heisenberg-transport equations, as well as in the
dynamical equations (\ref{4.14}), we consider time $t$ in a
coarse-grain temporal scale much longer than $|\Delta|^{-1}$ and
the coordinate $z$ on a coarse-grain spatial scale much longer
than $\lambdabar$. Any changes in the atomic subsystem and
displacement of atoms during the time increments comparable with
$|\Delta|^{-1}$ as well as any changes of the slow varying field
operators on the scale of a few $\lambdabar$ are ignored. We do
not actually know the exact behaviour neither for the atom nor for
the field Heisenberg operators, but we can approximately display
their averaged behaviour by solution of coupled equations
(\ref{4.14}) and (\ref{4.16}).

Instead of the Heisenberg-type equations for slowly varying
amplitudes $\hat{\epsilon}_i(z,t)$ and
$\hat{\epsilon}^\dagger_i(z,t)$ similar equations written for the
Stokes operators can be introduced
\begin{equation}
\left[\frac{\partial}{\partial z}+\frac{1}{c}%
\frac{\partial}{\partial t}\right]%
\hat{\Xi}_i(z,t)\;=\;%
\frac{i}{\hbar c}\left[\hat{\cal{H}}_{eff}(t),\,%
\hat{\Xi}_i(z,t)\right]%
\label{4.17}%
\end{equation}%
where $i=0,1,2,3$. All the terms appearing on the right hand sides
of equations (\ref{4.14}) and (\ref{4.17}) can be expressed via
the Stokes operators. Moreover it can be straightforwardly
verified that $\hat{\Xi}_0(z,t)$ stays unchanged as a function of
$t-z/c$, because of the conservation of the number of photons in
the forward scattering process. In turn this means that the first
term in Eq.(\ref{4.3}), i. e. the isotropic component of the
effective Hamiltonian, can be omitted in (\ref{4.17}) since it
commutes with all the Stokes operators.

As follows from the derived equations the main obstacle on the way
to convert the infinite number of operator equations to the system
of truncated equations, written for the collective atomic
variables, coupled with the integral or averaged field variables,
comes from the spatial dependence of the interaction process. Such
a dependence is caused by anisotropic terms in the effective
Hamiltonian, which lead to spatially varying entanglement of
different polarization modes of light with atomic spins along the
propagation path. The spatial profile of the Heisenberg-type
Stokes operators could be considered as uniform in $\Xi_2$
component and as accumulating the collective fluctuations of
atomic spins in complementary $\Xi_1$ component only in one
special case of a pure Faraday effect. In this case the
anisotropic components of the local susceptibilities of the medium
would be zero on average and would exist as fluctuations acting
only on the $\Xi_1$ Stokes component via random Faraday rotation.
This can be true if $\hat{T}_{10}^{(a)}(t)$,
$\hat{T}_{20}^{(a)}(t)$ and $\hat{T}_{2\pm 2}^{(a)}(t)$ operators
would be zero on average along the whole interaction cycle. But
even if we neglect the repopulation optical pumping mechanism,
coming from incoherent scattering, it would be not so easy to show
any realistic example of a proper atomic transition satisfying
this condition.

Such an example could be $j_0=1/2 \to j=1/2$ or $j_0=1/2 \to
j=3/2$ atomic transitions. Then there is no quadropole moment in
the ground state in principle and the ensemble consists of the
atoms perfectly polarized (oriented) in the direction orthogonal
to the propagation direction of the probe light. The atom with
spin one-half in its ground state does exist in reality (${}^3$He
for example), but it is not a convenient object for the
polarization sensitive experiments. In the earlier work on
entanglement and quantum information protocols with atomic
continuous variables \cite{K1,DCZP,K2,JKP,KzP1,Polzik} realistic
atoms were modelled as spin one half systems. One of the goals of
this paper is to analyze the applicability of such approximation.

Below we derive equations of motion in their general form and
discuss their possible conversion to the finite number of
wave-type equations written in terms of spatially distributed
collective variables.

\begin{figure}[tp]
\vspace{\baselineskip}
\includegraphics{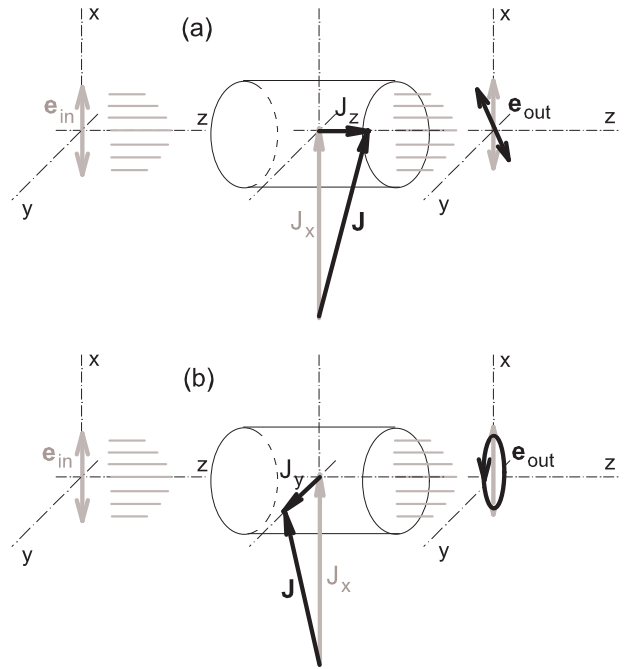}
\vspace{\baselineskip}%
\caption{Schematic diagram showing the polarization response in
the probe light, transmitted through the spin oriented atomic
sample, on random gyrotropy (a) and random birefringence (b). Both
processes are initiated by transverse fluctuations of the
collective angular momentum.}
\label{Fig.1}%
\end{figure}%

\section{Dynamics of the system in terms of collective
variables}\label{s5}

As the most important practical example, we will further discuss
the light propagation through atomic ensemble prepared originally
in the coherent spin state, see figure \ref{Fig.1}. We assume that
atoms fill the cylindrical volume with the cross section $S_0$
coinciding with the cross section of the probe light beam. The
atomic ensemble is located in the homogeneous magnetic field with
the direction orthogonal to the direction of the propagating
light. Originally atoms are perfectly oriented along the magnetic
field in the $x$ direction. Probe light is in a coherent state
linearly polarized along the $x$ direction. Such a geometry is
common,e.g., for experiments aiming at the quantum state
teleportation between field and atomic spin subsystems
\cite{DCZP,JKP,Polzik}.

\subsection{Field subsystem}

Let us consider first the evolution of the Stokes components.
Based on the commutation relation
\begin{equation}
\left[\hat{\Xi}_i(z',t), \hat{\Xi}_j(z,t)\right]\;=\;%
2i\varepsilon_{ijk}\,c\,\delta(z-z')\hat{\Xi}_k(z,t)%
\label{5.1}%
\end{equation}
where $\varepsilon_{ijk}=\pm 1$ in dependence on order of indices
$i\neq j\neq k$, equations (\ref{4.17}) can be rewritten as
follows
\begin{widetext}%
\begin{eqnarray}
\left[\frac{\partial}{\partial z}+\frac{1}{c}%
\frac{\partial}{\partial t}\right]%
\hat{\Xi}_1(z,t)&=&\phantom{+}%
\bar{\alpha}_1\sum_{a=1}^N\hat{T}_{10}^{(a)}(t)\,\delta(z_a(t)-z)\,%
\hat{\Xi}_3(z,t)\;-\;%
\bar{\alpha}_2\sum_{a=1}^N\hat{T}_{xy}^{(a)}(t)\,\delta(z_a(t)-z)\,%
\hat{\Xi}_2(z,t)%
\nonumber\\%
\left[\frac{\partial}{\partial z}+\frac{1}{c}%
\frac{\partial}{\partial t}\right]%
\hat{\Xi}_2(z,t)&=&\phantom{+}%
\bar{\alpha}_2\sum_{a=1}^N\hat{T}_{xy}^{(a)}(t)\,\delta(z_a(t)-z)\,%
\hat{\Xi}_1(z,t)\;-\;%
\bar{\alpha}_2\sum_{a=1}^N\hat{T}_{\xi\eta}^{(a)}(t)\,\delta(z_a(t)-z)\,%
\hat{\Xi}_3(z,t)%
\nonumber\\%
\left[\frac{\partial}{\partial z}+\frac{1}{c}%
\frac{\partial}{\partial t}\right]%
\hat{\Xi}_3(z,t)&=&%
-\bar{\alpha}_1\sum_{a=1}^N\hat{T}_{10}^{(a)}(t)\,\delta(z_a(t)-z)\,%
\hat{\Xi}_1(z,t)\;+\;%
\bar{\alpha}_2\sum_{a=1}^N\hat{T}_{\xi\eta}^{(a)}(t)\,\delta(z_a(t)-z)\,%
\hat{\Xi}_2(z,t)%
\label{5.2}%
\end{eqnarray}%
\end{widetext}%
where we introduced the dimensionless polarizabilities
\begin{equation}
\bar{\alpha}_i\;=\;\frac{4\pi\bar{\omega}}{S_0\,c}\,%
\alpha_i(\bar{\omega})\ \ \ \ \ i=1,2%
\label{5.2'}%
\end{equation}%
These equations are valid for any type of initial conditions and
can be simplified for excitation geometry described in the
preamble to this section.

The coherent forward scattering of light linearly polarized along
the $x$ direction does not modify the average angular momentum
orientation of the atomic ensemble. Transmitted light also
preserves its mean 100\% linear polarization. This means that in
equations (\ref{5.2}) only $\hat{T}_{xy}^{(a)}(t)$ and
$\hat{\Xi}_3(z,t)$ have non-zero expectation values. Other
operators exist only as fluctuating quantum variables. It is
possible to linearize these equations by substituting
$\hat{T}_{xy}^{(a)}(t)$ and $\hat{\Xi}_3(z,t)$ by their average
values and leaving in the right-hand side only the linearized
contribution over quantum fluctuations. Then the evolution of the
Stokes operators is given by
\begin{widetext}%
\begin{eqnarray}
\left[\frac{\partial}{\partial z}+\frac{1}{c}%
\frac{\partial}{\partial t}\right]%
\hat{\Xi}_1(z,t)&\approx&%
-\kappa_2\,\hat{\Xi}_2(z,t)\;+\;%
\bar{\alpha}_1\,\bar{\Xi}_3\,%
\sum_{a=1}^N\,\hat{T}_{10}^{(a)}(t)\,\delta(z_a(t)-z)\,%
\nonumber\\%
\left[\frac{\partial}{\partial z}+\frac{1}{c}%
\frac{\partial}{\partial t}\right]%
\hat{\Xi}_2(z,t)&\approx&\phantom{+}%
\kappa_2\,\hat{\Xi}_1(z,t)\;-\;%
\bar{\alpha}_2\,\bar{\Xi}_3\,%
\sum_{a=1}^N\,\hat{T}_{\xi\eta}^{(a)}(t)\,\delta(z_a(t)-z)\,%
\nonumber\\%
\left[\frac{\partial}{\partial z}+\frac{1}{c}%
\frac{\partial}{\partial t}\right]%
\hat{\Xi}_3(z,t)&\approx&\phantom{+}0%
\label{5.3}%
\end{eqnarray}%
\end{widetext}%
where $\bar{\Xi}_3=\bar{\Xi}_0$ is the average value of the
corresponding Stokes component, which is approximately unchanged
for the light beam propagating through the sample. The third line
in Eqs.(\ref{5.3}) just indicates this circumstance. The coupling
parameter
\begin{equation}
\kappa_2\;=\; \bar{\alpha}_2\,\sum_{a=1}^{N}\,%
\bar{T}_{xy}^{(a)}\,\delta(z_a(t)-z)%
\label{5.4}%
\end{equation}%
is responsible for birefringence effects, i.e. for unitary
transformation of linear polarization (defined in respect to
$\xi,\;\eta$ axes) to circular polarization and vice versa. Here
$\bar{T}_{xy}^{(a)}$ is the averaged and approximately unchanged
value of the alignment of $a$-th atom. For any atom, whose angular
momentum $j_0$ is oriented along the $x$ direction, the alignment
term is given by
\begin{equation}
\bar{T}_{xy}^{(a)}\;=\;%
\frac{[15j_0(2j_0-1)]^{1/2}}{2[2(j_0+1)(2j_0+1)(2j_0+3)]^{1/2}}\;%
\equiv\;\bar{T}_{xy}%
\label{5.5}%
\end{equation}%
The microscopic structure of $\kappa_2$ can be averaged over a
small mesoscopic interval of $\Delta z$ and substituted into
Eq.(\ref{5.3}) as $\kappa_2\,=\,\bar{\alpha}_2\bar{T}_{xy}n_0$,
where $n_0$ is the linear density of atoms (i. e. the number of
atoms per unit length). From the classical electrodynamics point
of view, the mesoscopically averaged product $\kappa_2\lambdabar$
is the difference between refraction indices for the polarizations
along $x$ and $y$ directions.

The most important terms in equations (\ref{5.3}) are the last
ones on the right hand sides. These terms show how any quantum
state, originally encoded in the spin fluctuations of the atomic
subsystem, can be mapped onto the polarization state of the light
subsystem. Thus these terms are responsible for the quantum
information processing in light - atoms interaction.

\subsection{Atomic subsystem}

Consider the dynamics of the orientation vector of an $a$-th atom.
Applying the commutation rule for irreducible tensor operators
\begin{eqnarray}
\left[\hat{T}_{KQ}^{(a)}, \hat{T}_{K'Q'}^{(b)}\right]&=&%
\delta_{ab}\,[(2K+1)(2K'+1)]^{1/2}\,%
\sum_{K''}%
\nonumber\\%
&&\times [1-(-)^{K+K'+K''}]\,%
\left\{\begin{array}{ccc}%
K & K' & K''\\ j_0 & j_0 & j_0%
\end{array}\right\}\,%
\nonumber\\%
&&\times\,(-)^{2j_0+K''}\,C^{K''Q''}_{KQ\, K'Q'}\,\hat{T}_{K''Q''}^{(a)}%
\label{5.6}%
\end{eqnarray}%
see Ref.\cite{VMK}, to the right hand side of equations
(\ref{4.14}) we obtain
\begin{widetext}%
\begin{eqnarray}
\dot{\hat{T}}{}_{10}^{(a)}(t)&=&%
\frac{i}{\hbar}\left[\hat{H}_0(t),\,\hat{T}_{10}^{(a)}(t)\right]\;-\;%
\bar{\alpha}_2\,(-)^{2j_0}\sqrt{10}\,%
\left\{\begin{array}{ccc}%
2 & 1 & 2\\ j_0 & j_0 & j_0%
\end{array}\right\}\,%
\hat{\Xi}_3(z_a(t),t)\,\hat{T}_{\xi\eta}^{(a)}(t)%
\nonumber\\%
&&+\;\bar{\alpha}_2\,(-)^{2j_0}\sqrt{10}\,%
\left\{\begin{array}{ccc}%
2 & 1 & 2\\ j_0 & j_0 & j_0%
\end{array}\right\}\,%
\hat{\Xi}_1(z_a(t),t)\,\hat{T}_{xy}^{(a)}(t),%
\nonumber\\%
\nonumber\\%
\dot{\hat{T}}{}_{1\pm 1}^{(a)}(t)&=&%
\frac{i}{\hbar}\left[\hat{H}_0(t),\,\hat{T}_{1\pm 1}^{(a)}(t)\right]\;\pm\;%
i\,\bar{\alpha}_2\,(-)^{2j_0}\frac{\sqrt{5}}{2}\,%
\left\{\begin{array}{ccc}%
2 & 1 & 2\\ j_0 & j_0 & j_0%
\end{array}\right\}\,%
\hat{\Xi}_0(z_a(t),t)\,\hat{T}_{2\pm 1}^{(a)}(t)%
\nonumber\\%
&&\mp\;i\,\bar{\alpha}_2\,(-)^{2j_0}\frac{\sqrt{5}}{2}\,%
\left\{\begin{array}{ccc}%
2 & 1 & 2\\ j_0 & j_0 & j_0%
\end{array}\right\}\,%
\hat{\Xi}_3(z_a(t),t)\,\hat{T}_{2\mp 1}^{(a)}(t)%
\nonumber\\%
&&+\;\bar{\alpha}_2\,(-)^{2j_0}\frac{\sqrt{5}}{2}\,%
\left\{\begin{array}{ccc}%
2 & 1 & 2\\ j_0 & j_0 & j_0%
\end{array}\right\}\,%
\hat{\Xi}_1(z_a(t),t)\,\hat{T}_{2\mp 1}^{(a)}(t)%
\nonumber\\%
&&\pm\;i\,\bar{\alpha}_1\,\frac{\sqrt{3}}%
{2[j_0(j_0+1)(2j_0+1)]^{1/2}}\;%
\hat{\Xi}_2(z_a(t),t)\,\hat{T}_{1\pm 1}^{(a)}(t)%
\label{5.7}%
\end{eqnarray}%
\end{widetext}%
As one can see these equations are not closed, since the right
hand sides are expressed in terms of alignment components.
Moreover the higher rank multipoles drive the dynamics of the
alignment components. Thus to introduce the system of closed
Heisenberg equations it is necessary to consider the coupled
dynamics of all irreducible components defined for each atom.

However the dynamics of the atomic subsystem can be approximated
by the dynamics of orientation components only as far as we are
restricted to the discussion of the excitation regime described in
the preamble to this section. As shown in the appendix \ref{a}
equations (\ref{5.7}) can be transformed to the set of nonlinear
equations, where the right-hand sides are expressed in terms of
the operators of the angular momentum. In a special, but the most
important for us, regime of small fluctuations such transformed
equations can be linearized and simplified to the equations
describing the dynamics of the vector of the angular momentum in a
closed form. Let us define the vector of the collective angular
momentum of ensemble as
\begin{equation}
\hat{\mathbf J}(t)\;=\;\sum_{a=1}^{N}\,\hat{\mathbf j}^{(a)}(t)%
\label{5.8}%
\end{equation}%
Then starting from equations (\ref{a.7}), written for any $a$-th
atom, and making the sum over all atoms of the ensemble, we arrive
at the following equations governing the dynamics of the
collective angular momentum
\begin{eqnarray}
\dot{\hat{J}}_z(t)&\approx&\phantom{+}%
(\Omega_0\,+\,\Omega_2)\,\hat{J}_y(t)\;-\;%
\bar{T}_{xy}\,\bar{\alpha}_2\,%
\sum_{a=1}^{N}\,\hat{\Xi}_1(z_a(t),t)%
\nonumber\\%
\dot{\hat{J}}_y(t)&\approx&-(\Omega_0\,+\,\Omega_2)\,\hat{J}_z(t)\;+\;%
\frac{1}{2}\bar{T}_{x}\,\bar{\alpha}_1\,%
\sum_{a=1}^{N}\,\hat{\Xi}_2(z_a(t),t)%
\nonumber\\%
\dot{\hat{J}}_x(t)&\approx&\phantom{+}0
\label{5.9}%
\end{eqnarray}%
where parameters $\Omega_2$ and $\bar{T}_{x}$ are defined by
expressions (\ref{a.8}) and (\ref{a.9}) respectively and
$\bar{T}_{xy}$ is given by (\ref{5.5}). The last source-type terms
on the right hand side of these equations are responsible for the
mapping of any quantum state, originally prepared in the Stokes
fluctuations of the field subsystem, onto long lived atomic spin
subsystem.

Equations (\ref{5.3}) and (\ref{5.9}), considered together,
approximate the dynamical evolution of coupled collective
variables of light and atomic subsystems. But as follows from the
structure of these equations they are still not closed because the
coupling terms {\it are not expressed by collective variables}. As
we see from (\ref{5.3}) only the atoms currently located before
the wavefront contribute into formation of the fluctuating Stokes
variables. In turn, such spatially dependent Stokes fluctuations,
which stay actually unknown, drive the dynamics of atomic
collective angular momentum via source terms in (\ref{5.9}).
However, as we show below, under certain simplifying assumptions,
these equations can be further transformed into the system of
closed equations describing the wave-type spatial and temporal
distribution of the collective Heisenberg operators of the atomic
and field subsystems.

\subsection{Mesoscopic averaging}
If atoms are slowly drifting in space and during the interaction
with a short probe light pulse each atom preserves its location
inside the area much less than the length scale comparable with
$\kappa_2^{-1}$ or with the sample size, equations (\ref{5.3}) and
(\ref{5.9}) can be transformed into a closed form. Let us note
here that if in the experiment the duration of a probe pulse is
chosen less than a few microseconds such a condition is normally
fulfilled even in case of atoms at room temperature. For the case
of cold trapped atoms this assumption is consistent with the pulse
duration up to a second. Then the equations of motion can be
rewritten for any mesoscopic layer of the sample, which gives only
a small increment to the Heisenberg operators but contains a large
number of atoms. The atoms do not leave the layer during the
interaction time and cooperatively interact with the
electromagnetic field. If this layer has a length of $\Delta z$,
we can introduce the averaged Stokes operators $\hat{\Xi}_1(z,t)$
along this layer. As a next step, instead of total angular
momentum, given by Eq.(\ref{5.8}) we can define its mesoscopic
spatial distribution as follows
\begin{equation}
\hat{\cal\mathbf J}(z,t)\;=\;\frac{1}{\Delta z}%
\sum_{z<z_a<z+\Delta z}\,%
\hat{\mathbf j}^{(a)}(t)\,%
\label{5.10}%
\end{equation}%
where the sum over $a$ is extended only over the atoms located
inside the layer. Then the total angular momentum of the ensemble
is expressed as
\begin{equation}
\hat{\mathbf J}(t)\;=\;\int_{0}^{L}%
\hat{\cal\mathbf J}(z,t)\,dz%
\label{5.11}%
\end{equation}%
where $dz=\Delta z$ and $L$ is the length of the sample.

The above assumptions lead us to the following set of closed and
coupled equations for the mesoscopically averaged spatial
distributions of the field and atomic variables
\begin{eqnarray}
\left[\frac{\partial}{\partial z}+\frac{1}{c}%
\frac{\partial}{\partial t}\right]%
\hat{\Xi}_1(z,t)&=&%
-\kappa_2\,\hat{\Xi}_2(z,t)\;+\;%
2\beta\,\bar{\Xi}_3\,%
\hat{{\cal J}}_z(z,t)%
\nonumber\\%
\left[\frac{\partial}{\partial z}+\frac{1}{c}%
\frac{\partial}{\partial t}\right]%
\hat{\Xi}_2(z,t)&=&%
\phantom{+}\kappa_2\,\hat{\Xi}_1(z,t)\;-\;%
2\epsilon\,\bar{\Xi}_3\,%
\hat{{\cal J}}_y(z,t)%
\nonumber\\%
\frac{\partial}{\partial t}\hat{{\cal J}}_z(z,t)&=&%
\phantom{+}\Omega\,\hat{{\cal J}}_y(z,t)\;-\;%
\theta_y\bar{{\cal J}}_x\,%
\hat{\Xi}_1(z,t)%
\nonumber\\%
\frac{\partial}{\partial t}\hat{{\cal J}}_y(z,t)&=&%
-\Omega\,\hat{{\cal J}}_z(z,t)\;+\;%
\theta_z\bar{{\cal J}}_x\,%
\hat{\Xi}_2(z,t)%
\nonumber\\%
&&\label{5.12}%
\end{eqnarray}%
where we used the same notation for the averaged Stokes variables
as for their microscopic origins and denoted the  mesoscopic
spatial distributions of angular momentum components as $\hat{\cal
J}_{\mu}(z,t)$, with $\mu=x,y,z$. It is taken into account that
$\bar{{\cal J}}_x(z,t)=\bar{{\cal J}}_x=const$. Equations
(\ref{5.12}) should be accompanied by corresponding initial and
boundary conditions, which are given by
\begin{eqnarray}
\hat{\Xi}_1(0,t)&=&\hat{\Xi}_1^{in}(t)%
\nonumber\\%
\hat{\Xi}_2(0,t)&=&\hat{\Xi}_2^{in}(t)%
\nonumber\\%
\hat{{\cal J}}_y(z,0)&=&\hat{{\cal J}}_y^{in}(z)%
\nonumber\\%
\hat{{\cal J}}_z(z,0)&=&\hat{{\cal J}}_z^{in}(z)%
\label{5.13}%
\end{eqnarray}%
The solution of these equations presented in the next section
shows how the swapping of quantum fluctuations between light and
spin subsystems takes place during the interaction process.

Several new parameters appear in equations (\ref{5.12}). Firstly,
by $\Omega$ we denote the frequency $\Omega=\Omega_0+\Omega_2$ of
the regular precession caused by the external magnetic field as
well as by the light-induced shift of the Zeeman sublevels.
Secondly, there are two new parameters in the first pair of
equations describing the transformation of the Stokes variables.
The angle $\beta$
\begin{equation}
\beta\;=\;\frac{\sqrt{3}}{2[j_0(j_0+1)(2j_0+1)]^{1/2}}\,%
\bar{\alpha}_1%
\label{5.14}%
\end{equation}%
is the angle of Faraday-type rotation of the polarization plane of
the propagating light per one spin flip in the ensemble in
$z$-direction. The parameter $\epsilon$
\begin{equation}
\epsilon\;=\;\frac{[15(2j_0-1)]^{1/2}}%
{2[2j_0(j_0+1)(2j_0+1)(2j_0+3)]^{1/2}}\,\bar{\alpha}_2%
\label{5.15}%
\end{equation}%
is the ellipticity induced in the propagating light by the atomic
sample per one spin flip in $y$-direction, see figure \ref{Fig.1}.
Thirdly, there are two angles in the second pair of equations
describing the dynamics of the spatial distribution of the atomic
angular momenta. Angle $\theta_y=\epsilon$ is the rotation angle
of the local collective angular momentum, originally oriented
along $x$-axis, around $y$-axis per one photon propagating trough
the sample in either $\xi$-type or $\eta$-type linear
polarization. In turn, angle $\theta_z=\beta$ is the rotation
angle of the local angular momentum around $z$-axis per one photon
propagating through the sample in either right-hand-type or
left-hand-type circular polarization.

\section{Entanglement of the quantum states of light and
atoms}\label{s6}

For pedagogical purposes, we consider at first a special example
of optical excitation in the far-off-resonance wing of $j_0=1/2\to
j=1/2$ or $j_0=1/2\to j=3/2$ optical transitions and will discuss
the general case after that.

\subsection{Example of $j_0=1/2\to j=1/2,\,j=3/2$ optical
transitions}\label{s6.a}

In the case of $j_0=1/2\to j=1/2$ or $j_0=1/2\to j=3/2$ transition
equations (\ref{5.12}) are simplified to the following form
\begin{eqnarray}
\left[\frac{\partial}{\partial z}\,+\,\frac{1}{c}%
\frac{\partial}{\partial t}\right]\hat{\Xi}_1(z,t)&=&%
\phantom{+}2\beta\,\bar{\Xi}_3\,%
\hat{{\cal J}}_z(z,t)%
\nonumber\\%
\left[\frac{\partial}{\partial z}\,+\,\frac{1}{c}%
\frac{\partial}{\partial t}\right]\hat{\Xi}_2(z,t)&=&\phantom{+}0%
\nonumber\\%
\frac{\partial}{\partial t}\hat{{\cal J}}_z(z,t)&=&%
\phantom{+}\Omega_0\,\hat{{\cal J}}_y(z,t)%
\nonumber\\%
\frac{\partial}{\partial t}\hat{{\cal J}}_y(z,t)&=&%
-\Omega_0\,\hat{{\cal J}}_z(z,t)\;+\;%
\beta\bar{{\cal J}}_x\,%
\hat{\Xi}_2(z,t)%
\nonumber\\%
&&\label{6.1}%
\end{eqnarray}%
where it was taken into account that $\bar{\alpha}_2=0$ and
therefore $\kappa_2=0$, $\Omega_2=0$, $\epsilon=\theta_y=0$.
Because of the second line in the system (\ref{6.1}), one has
$\hat{\Xi}_2(z,t)=\hat{\Xi}_2^{in}(t-z/c)$.

The straightforward solution for the distribution of the angular
momentum components leads to
\begin{widetext}%
\begin{eqnarray}
\hat{{\cal J}}_z(z,t)&=&%
\phantom{+}\cos\Omega_0t\;\hat{{\cal J}}_z^{in}(z)\,+\,%
\sin\Omega_0t\;\hat{{\cal J}}_y^{in}(z)\;+\;%
\beta\bar{{\cal J}}_x\,%
\int_0^t dt'\sin\Omega_0(t-t')\;%
\hat{\Xi}_2^{in}(t'-z/c)%
\nonumber\\%
\hat{{\cal J}}_y(z,t)&=&%
-\sin\Omega_0t\;\hat{{\cal J}}_z^{in}(z)\,+\,%
\cos\Omega_0t\;\hat{{\cal J}}_y^{in}(z)\,+\,%
\beta\bar{{\cal J}}_x\;\int_0^t dt'\cos\Omega_0(t-t')\;%
\hat{\Xi}_2^{in}(t'-z/c)%
\label{6.2}%
\end{eqnarray}%
and the Stokes component $\hat{\Xi}_1(z,t)$ is given by
\begin{eqnarray}
\hat{\Xi}_1(z,t)&=&\hat{\Xi}_1^{in}(t-z/c)\,+\,%
2\beta\,\bar{\Xi}_3\,\int_0^z \!dz'\,%
\hat{{\cal J}}_z(z',t-(z-z')/c)\;=\;\hat{\Xi}_1^{in}(t-z/c)%
\nonumber\\%
&&+\;2\beta\,\bar{\Xi}_3\,%
\int_0^z \!dz'\,\cos\Omega_0[t-(z-z')/c]\;%
\hat{{\cal J}}_z^{in}(z')%
\;+\;2\beta\,\bar{\Xi}_3\,%
\int_0^z \!dz'\,\sin\Omega_0[t-(z-z')/c]\;%
\hat{{\cal J}}_y^{in}(z')%
\nonumber\\%
&&+\;2\beta^2\,\bar{\Xi}_3\bar{{\cal J}}_x%
\int_0^z \!dz'\!\int_0^{t-(z-z')/c}\!dt'\sin\Omega_0[t-t'-(z-z')/c]\;%
\hat{\Xi}_2^{in}(t'-z'/c)%
\label{6.3}%
\end{eqnarray}%
\end{widetext}%
This solution explains the basic idea of entanglement of the
collective quantum states of atomic and field subsystems. After
the interaction cycle the quantum state of light described by the
$\hat{\Xi}_2^{in}$ component is mapped onto the atomic angular
momentum components because of the last terms in Eqs.(\ref{6.2}).
In turn, the information written in either $\hat{{\cal J}}_z^{in}$
or $\hat{{\cal J}}_y^{in}$ components of the angular momentum is
mapped onto the $\hat{\Xi}_1$ Stokes component of light, because
of the second and third terms in Eq.(\ref{6.3}). Moreover, due to
atom-field interaction there is a partial mapping of the
$\hat{\Xi}_2^{in}$ component onto $\hat{\Xi}_1$ component because
of the last term in Eq.(\ref{6.3}).

The retardation effects, which are clearly visible in the derived
solution are mainly important in order to preserve the proper
commutation relations for the whole set of the Heisenberg
operators involved in the process. The initial zero moment of time
is a conventional and arbitrary chosen parameter to coordinate
Schr\"{o}dinger and Heisenberg pictures. Physically the
interaction cycle starts only when the wavefront of the probe
radiation crosses $z=0$ point. We will take this time as the
initial moment $t=0$. In the Heisenberg formalism this means that
any expectations values of $\hat{\Xi}_1^{in}(t)$ and
$\hat{\Xi}_2^{in}(t)$ operators or their products are equal to
zero for time $t<0$. Since the goal is to evaluate the expectation
values of outgoing operators it is acceptable to substitute the
lower zero limit by $t\to z/c$ in (\ref{6.2}) and by $t'\to z'/c$
in the internal integral (\ref{6.3}). Then the derived solution
can be further applied at any spatial point $z$ only for time
$t>z/c$, which is in accordance with physical manifestation of
retardation effects.

In reality, we are most interested in a correct description of the
behavior of the modes around the carrier frequency $\bar{\omega}$,
see definition (\ref{4.15}). For the modes distributed in the
spectral domain much narrower than $c/L$, where $L$ is the length
of the sample, the retardation effects become negligible. In this
case the above solution is simplified and can be written in terms
of transverse components of the collective angular momentum
\begin{widetext}%
\begin{eqnarray}
\hat{J}_z(t)&=&%
\phantom{+}\cos\Omega_0t\;\hat{J}_z^{in}\,+\,%
\sin\Omega_0t\;\hat{J}_y^{in}\;+\;%
\beta\bar{J}_x\,%
\int_0^t dt'\sin\Omega_0(t-t')\;%
\hat{\Xi}_2^{in}(t')%
\nonumber\\%
\hat{J}_y(t)&=&%
-\sin\Omega_0t\;\hat{J}_z^{in}\,+\,%
\cos\Omega_0t\;\hat{J}_y^{in}\,+\,%
\beta\bar{J}_x\;\int_0^t dt'\cos\Omega_0(t-t')\;%
\hat{\Xi}_2^{in}(t')%
\label{6.4}%
\end{eqnarray}%
\end{widetext}%
and for the output Stokes operator $\hat{\Xi}_1^{out}(t)$
\begin{eqnarray}
\hat{\Xi}_1^{out}(t)&=&\hat{\Xi}_1^{in}(t)%
\phantom{+\;2\beta\,\bar{\Xi}_3\cos\Omega_0t\,\hat{J}_z^{in}\;+\;%
2\beta\,\bar{\Xi}_3\sin\Omega_0t\,\hat{J}_y^{in}}%
\nonumber\\%
&&+\;2\beta\,\bar{\Xi}_3\cos\Omega_0t\,\hat{J}_z^{in}\;+\;%
2\beta\,\bar{\Xi}_3\sin\Omega_0t\,\hat{J}_y^{in}%
\nonumber\\%
&&+\;2\beta^2\,\bar{\Xi}_3\bar{J}_x\,%
\int_0^{t} dt'\sin\Omega_0(t-t')\;%
\hat{\Xi}_2^{in}(t')%
\label{6.5}%
\end{eqnarray}%
which coincides, in principle, with the solution obtained earlier
in Refs.\cite{JKP,H}.

The input-output transformations expressed by (\ref{6.4}) and
(\ref{6.5}) have a simple logical structure and therefore are
attractive for possible applications to the quantum information
tasks. However, they are only valid for $j_0=1/2\to j=1/2,j=3/2$
isolated transitions, and it is rather difficult to find a
practical example of such a transition. However, the same simple
input-output relations are effectively realized in the case of
alkali atoms excited in the far wing of $D_1$ or $D_2$ lines. In
this case the contributions from the components of the upper
hyperfine multiplet to the interaction add up in such a way that
the alignment effects become unimportant. For a transition from
the ground hyperfine state $F_0$ to all hyperfine states $F$ we
straightforwardly obtain after taking the sum over the excited
states
\begin{eqnarray}
\sum_{F}\,\bar{\alpha}_2&\propto&%
\sum_{F}\,\frac{|d_{j_0j}|^2}{-\hbar\Delta_F}\,(-)^{F+F_0+1}\,(2F+1)(2F_0+1)\,%
\nonumber\\%
&&\times\left\{\begin{array}{ccc}1&1&2\\F_0&F_0&F\end{array}\right\}\,%
\left\{\begin{array}{ccc}I&j&F\\1&F_0&j_0\end{array}\right\}^{2}
\;\to\; 0%
\label{6.6}%
\end{eqnarray}%
Hence the alignment contribution vanishes for the ground state
electronic angular momentum $j_0=1/2$ if the frequency detuning
$\Delta_F=\bar{\omega}-\omega_{FF_0}$ for each hyperfine
transition $F_0\to F$ becomes larger than the hyperfine splitting
of the upper state. The crucial inequality $|\bar{\alpha}_1|\gg
|\bar{\alpha}_2|$ which allows to neglect the alignment effects is
fulfilled the better the less important is the dependence on the
upper hyperfine momentum in the denominator of (\ref{6.6}). In
spite of the fact that for heavy alkali atoms the ratio of the
hyperfine to fine splitting in the upper state $\delta
E_{hf}/\delta E_f$ is really small (it is less than $10^{-4}$ for
${}^{133}$Cs), which makes it possible to neglect the alignment
contribution for large detunings, such an approximation is not
always easy to fulfill under typical experimental conditions. This
is because for a large frequency detuning one has to use pulses
with a very large photon number in order to make the atom-field
interaction strong enough. Large photon numbers are often
inconvenient due to either limitations of detectors or due to a
limited pulse duration, or both. Therefore the frequency detuning
$\Delta$ is frequently chosen to be comparable with $\delta
E_{hf}$. A typical detuning chosen in the experiment \cite{JKP},
is of the order of $900\; MHz$ with the hyperfine splitting in
$6{}^{2}P_{3/2}$ of $\delta E_{hf}\sim 200\; MHz$. This is one of
the motivations for the deeper analysis of equations (\ref{5.12})
in their general form.

\subsection{General solution}

Equations (\ref{5.12}) can be solved in the general case, which we
do first ignoring the retardation effects. This is the most
important case for practical applications to quantum information
processing. Without retardation the system of coupled equations
(\ref{5.12}) can be rewritten as follows
\begin{eqnarray}
\frac{\partial}{\partial z}%
\hat{\Xi}_1(z,t)&=&%
-\kappa_2\,\hat{\Xi}_2(z,t)\;+\;%
2\beta\,\bar{\Xi}_3\,%
\hat{{\cal J}}_z(z,t)%
\nonumber\\%
\frac{\partial}{\partial z}%
\hat{\Xi}_2(z,t)&=&%
\phantom{+}\kappa_2\,\hat{\Xi}_1(z,t)\;-\;%
2\epsilon\,\bar{\Xi}_3\,%
\hat{{\cal J}}_y(z,t)%
\nonumber\\%
\frac{\partial}{\partial t}\hat{{\cal J}}_z(z,t)&=&%
\phantom{+}\Omega\,\hat{{\cal J}}_y(z,t)\;-\;%
\theta_y\bar{{\cal J}}_x\,%
\hat{\Xi}_1(z,t)%
\nonumber\\%
\frac{\partial}{\partial t}\hat{{\cal J}}_y(z,t)&=&%
-\Omega\,\hat{{\cal J}}_z(z,t)\;+\;%
\theta_z\bar{{\cal J}}_x\,%
\hat{\Xi}_2(z,t)%
\label{6.7}%
\end{eqnarray}%
This is a set of differential equations with constant coefficients
and it is accompanied by initial and boundary conditions
(\ref{5.13}). The solution can be presented in the form of an
integral response on these conditions. For the Stokes operators it
can be written as
\begin{eqnarray}
\hat{\Xi}_i(z,t)&=&%
\sum_{j=1,2}\,\int_0^t dt'\,%
{\cal M}_{ij}(z,t-t')\,\hat{\Xi}_j^{in}(t')%
\phantom{{\cal F}_{i\nu}(z-z',t)\,}%
\nonumber\\%
&&+\;\sum_{\nu=z,y}\,\int_0^z dz'\,%
{\cal F}_{i\nu}(z-z',t)\,\hat{{\cal J}}_{\nu}^{in}(z')%
\label{6.8}%
\end{eqnarray}%
with $i=1,2$. For the operators of the spatial distributions of
the angular momentum components the solution is
\begin{eqnarray}
\hat{{\cal J}}_{\mu}(z,t)&=&%
\sum_{j=1,2}\,\int_0^t dt'\,%
{\cal G}_{\mu j}(z,t-t')\,\hat{\Xi}_j^{in}(t')%
\phantom{{\cal N}_{\mu\nu}(z-z',t)\,}%
\nonumber\\%
&&+\;\sum_{\nu=z,y}\,\int_0^z dz'\,%
{\cal N}_{\mu\nu}(z-z',t)\,\hat{{\cal J}}_{\nu}^{in}(z')%
\label{6.9}%
\end{eqnarray}%
with $\mu=z,y$. The solution can be formally extended over the
space-time region $0<t<\infty$ and $0<z<\infty$.

As shown in the appendix \ref{b}, the kernels of the integral
transformations in (\ref{6.8}) and (\ref{6.9}) can be expressed
via the corresponding Laplace images
\begin{equation}
{\cal K}(z,t)\;=\;\frac{1}{(2\pi i)^2}%
\int_{p_0-i\infty}^{p_0+i\infty}%
\int_{s_0-i\infty}^{s_0+i\infty}%
{\cal K}(p,s)\,e^{pz+st}\,dp\,ds%
\label{6.10}%
\end{equation}%
where ${\cal K}(z,t)$ is any of the matrix functions ${\cal
M}(z,t),\ldots,{\cal N}(z,t)$ in equations (\ref{6.8}) and
(\ref{6.9}) and ${\cal K}(p,s)$ is its Laplace image. The limits
of integration $p_0$ and $s_0$ are chosen as arbitrary real values
warranting the existence of Laplace images for the spatial and
temporal transforms respectively. In the appendix \ref{b} the
images for all the matrices ${\cal M}(p,s),\ldots,{\cal N}(p,s)$
are calculated explicitly. Evaluation of the integrals
(\ref{6.10}) can be done numerically for any sample when all the
external parameters are defined.

Let us briefly explain how the retardation effects can be taken
into account. To do this the retardation time $\tau=t-z/c$ should
be introduced as an independent variable for each spatial point
$z$ at a certain moment of time $t$. Then the basic system
(\ref{5.12}), where all the operators are considered now as
functions of $z$ and $\tau$, can be converted to the form
(\ref{6.7}) but with the derivative over retardation time on the
left hand side. The initial conditions should be also modified and
written for the initial retardation time $\tau=0$, which
corresponds now to spatially dependent moments in the laboratory
time $t=z/c$. As before we will assume that the wavefront of the
probe radiation crosses $z=0$ point at zero time. In this case we
need to know the cooperative atom-field dynamics at any spatial
point $z$ only for time $t>z/c$. Then we may ignore the
interaction Hamiltonian for time $0<t<z/c$ and chose the initial
conditions in (\ref{5.13}) in the following form
\begin{eqnarray}
\hat{{\cal J}}_z(z,\tau=0)&=&%
\phantom{+}\cos(\Omega_0z/c)\;\hat{{\cal J}}_z^{in}(z)%
\phantom{\sin(\Omega_0z/c)\;\hat{{\cal J}}_y^{in}(z)}%
\nonumber\\%
&&+\,\sin(\Omega_0z/c)\;\hat{{\cal J}}_y^{in}(z)%
\;\equiv\;\hat{{\cal J}}_z^{in\prime}(z)%
\nonumber\\%
\hat{{\cal J}}_y(z,\tau=0)&=&%
-\sin(\Omega_0z/c)\;\hat{{\cal J}}_z^{in}(z)%
\nonumber\\%
&&+\,\cos(\Omega_0z/c)\;\hat{{\cal J}}_y^{in}(z)%
\;\equiv\;\hat{{\cal J}}_y^{in\prime}(z)%
\label{6.11}%
\end{eqnarray}%
There are no changes in the boundary conditions in (\ref{5.13}),
since at $z=0$ the retardation time $\tau$ coincides with $t$.
Thus in the most general case the solution
(\ref{6.8})-(\ref{6.10}) should be rewritten in terms of the
retardation time and should depend on $\hat{{\cal
J}}_z^{in\prime}(z)$ and $\hat{{\cal J}}_y^{in\prime}(z)$ as on
initial quantum fluctuations of the atomic angular momenta.

Finally the retardation effects give us the following correction
to the spatial and temporal dynamics of the field variables
\begin{widetext}%
\begin{equation}
\hat{\Xi}_i(z,t)\;=\;%
\sum_{j=1,2}\,\int_0^{t-z/c} dt'\,%
{\cal M}_{ij}(z,t-z/c-t')\,\hat{\Xi}_j^{in}(t')\;+\;%
\sum_{\nu=z,y}\,\int_0^z dz'\,%
{\cal F}_{i\nu}(z-z',t-z/c)\,\hat{{\cal J}}_{\nu}^{in\prime}(z')%
\label{6.12}%
\end{equation}%
and of the atomic variables
\begin{equation}
\hat{{\cal J}}_{\mu}(z,t)\;=\;%
\sum_{j=1,2}\,\int_0^{t-z/c} dt'\,%
{\cal G}_{\mu j}(z,t-z/c-t')\,\hat{\Xi}_j^{in}(t')\;+\;%
\sum_{\nu=z,y}\,\int_0^z dz'\,%
{\cal N}_{\mu\nu}(z-z',t-z/c)\,\hat{{\cal J}}_{\nu}^{in\prime}(z')%
\label{6.13}%
\end{equation}%
\end{widetext}%
These expressions can be applied for calculation of expectation
values of any products of operators at a spatial point $z$ for the
time $t>z/c$.

An important feature of solutions (\ref{6.12}) and (\ref{6.13}) is
their wave nature. As follows from the Laplace transform carried
out in the Appendix \ref{b} the eigenmodes of spatial and temporal
distributions of atoms-field collective fluctuations are
associated with the poles of determinant $\Delta(p,s)\to 0$, where
$\Delta(p,s)$ is defined by (\ref{b.3}). For atoms with spin $1/2$
this condition leads to $p\to 0$ and $s\to \pm{\rm i}\Omega_0$. In
the absence of the magnetic field $\Omega_0\to 0$ there are only
two collective modes associated with the total number of atoms and
with the total number of scattering photons which undergo
polarization sensitive interaction in the process of coherent
forward scattering. This makes it possible to further simplify the
effective Hamiltonian and to discuss it in terms of integrated
collective variables of the canonical type \cite{KzPl,JKP}.

However in a general case, due to presence of alignment associated
effects, the condition $\Delta(p,s)\to 0$ gives coupled roots
$s=s(p)$, i.e the wave-type modes which can survive in the sample
and be excited by either time dependent fluctuations of the field
Stokes components or by spatial fluctuations of atomic spins. Such
quantum superposition of the field and material waves can be
understood as a polariton-type spin wave generated in the medium.
The multimode structure of such waves does not allow entanglement
only between integrated collective variables. Instead, in general
the entanglement is distributed in all polariton modes. From the
quantum information point of view this means that in the case of
atoms with the ground state angular momentum equal or larger than
$1$ a possible entanglement resource hidden in the polariton modes
allows for multimode entanglement rather than a single mode
entanglement available for spin one half systems. The quantum
information then will be mapped on spatially dependent
correlations of an atomic standing spin wave.

The following remark concerning the terminology is due here. Let
us point out the principle difference between the spin-wave
dynamics described by Eqs.(5.13) and the collective spin behavior
in the spin-organized cold fermion gas, which was discussed years
ago, see \cite{BLL}. In our case the cooperative dynamics of
atomic spins is driven by the radiation-type interaction whereas
in the cold fermionic systems it is driven by the static
interparticle interaction via the longitudinal electric field.

\subsection{Spin squeezing in an ensemble of
cesium atoms.}\label{s6.c}

In this section we concentrate on calculations relevant for the
existing experimental example \cite{JKP} carried out with
ensemble(s) of ${}^{133}$Cs atoms. We first calculate the degree
of spin squeezing ignoring the alignment, i.e., following the
model adopted in \cite{JKP} and describing the atomic ground state
by the spin (orientation) only. We then present the numerical
results including the alignment and show how this affects spin
squeezing of atoms.

In the experiment \cite{JKP} the entanglement (two mode squeezing)
was generated between two spatially separated via a Faraday-type
detection of light. For pedagogical reasons we will make numerical
simulations for the case of single mode spin squeezing, which
makes no difference for the present discussion. We will completely
ignore the retardation effects and consider the spin dynamics
without external magnetic field.

In the case of cesium atoms the alignment contribution can be
suppressed if the frequency detuning of the probe light from the
atomic resonance is much larger than the hyperfine splitting in
the upper state, as described in the end of section \ref{s6.a}. If
the alignment is ignored the basic equations stem from the
expressions (\ref{6.4}) and (\ref{6.5}), which in the absence of
magnetic field read
\begin{eqnarray}
\hat{J}_z(t)&=&%
\phantom{+}\hat{J}_z^{in}%
\nonumber\\%
\hat{J}_y(t)&=&\hat{J}_y^{in}\,+\,%
\beta\bar{J}_x\;\int_0^t dt'\hat{\Xi}_2^{in}(t')%
\nonumber\\%
\hat{\Xi}_1^{out}(t)&=&\hat{\Xi}_1^{in}(t)\;+\;%
2\beta\,\bar{\Xi}_3\,\hat{J}_z^{in}%
\nonumber\\%
\hat{\Xi}_2^{out}(t)&=&\hat{\Xi}_2^{in}(t)%
\label{6.14}%
\end{eqnarray}%
 The input-output transformations (\ref{6.14}) show the entangling
mechanism of the atoms-field variables in the process of coherent
forward scattering. If the number of atoms and photons is large
enough such as $\beta \bar{J}_x\gg 1$ and $\beta\bar{\Xi}_3t\gg 1$
the output quantum fluctuations become strongly entangled. Indeed,
in this case the role of $\hat{J}_y^{in}$ and
$\hat{\Xi}_1^{in}(t)$ terms on the right hand side of (\ref{6.14})
becomes negligible if these fluctuations have been originally
Poissonian. After the interaction there would be strong
correlations between fluctuations of $\int_0^t
\hat{\Xi}_1^{out}(t')dt'$ and $\hat{J}_z(t)$ as well as between
$\hat{J}_y(t)$ and $\int_0^t \hat{\Xi}_2^{out}(t')dt'$. If the
observable $\int_0^t \hat{\Xi}_1^{out}(t')dt'$ is measured by a
balanced Faraday detector there will be no longer standard quantum
uncertainty in $\hat{J}_z(t+0)$, whose original (at $t\to 0$)
Poissonian fluctuations will be suppressed by a factor
$\beta\bar{\Xi}_3t\gg 1$. The collective spin state will become
squeezed and will accumulate its quantum uncertainty in
$\hat{J}_y(t+0)$ fluctuation.

It is important to recognize that even in such an ideal scheme
there is a restriction on the number of participating atoms and
photons because of accumulation of events of the incoherent
scattering. It is clear that in each event of the incoherent
scattering one photon and one atom of the ensemble cancel out from
entangling process. These losses can be neglected if the number of
scattering events is much less than the total numbers of atoms
$N_{a}$ and incoming photons $N_{ph}$. This can be written as two
inequalities
\begin{eqnarray}
\frac{\sigma_{\Delta}}{S_0}\,N_{ph}N_{a}&\ll& N_{a}%
\nonumber\\%
\frac{\sigma_{\Delta}}{S_0}\,N_{ph}N_{a}&\ll& N_{ph}%
\label{6.15}%
\end{eqnarray}%
where $\sigma_{\Delta}$ is the cross section for off-resonant
incoherent scattering in full solid angle, and $S_0$ is the area
of the light beam, which coincides with the area of atomic cloud.
Both inequalities lead to similar restrictions on the number of
participating atoms and photons
\begin{equation}
N_{a},N_{ph}\;\ll\;\frac{S_0}{\sigma_{\Delta}}%
\label{6.16}%
\end{equation}%
In case of photons this inequality can be also understood as a
restriction on the whole interaction time $T$ (the probe pulse
duration), since $N_{ph}=\bar{\Xi}_0T= \bar{\Xi}_3T$. Since both
type of the losses are undesirable one can assume that
$N_{a}=N_{ph}$ to exclude any preference for atoms or photons. The
number of scattered photons can be expressed as
\begin{equation}
N_{ph}\;=\;\eta \frac{S_0}{\sigma_{\Delta}}%
\label{6.17}%
\end{equation}%
where by small parameter $\eta$ we denoted the relative number of
atoms and photons lost as a result of incoherent scattering, see
the first line of (\ref{6.15}). A detailed analysis of the role of
spontaneous emission is presented in \cite{Polzik}.

\begin{figure}[tp]
\includegraphics{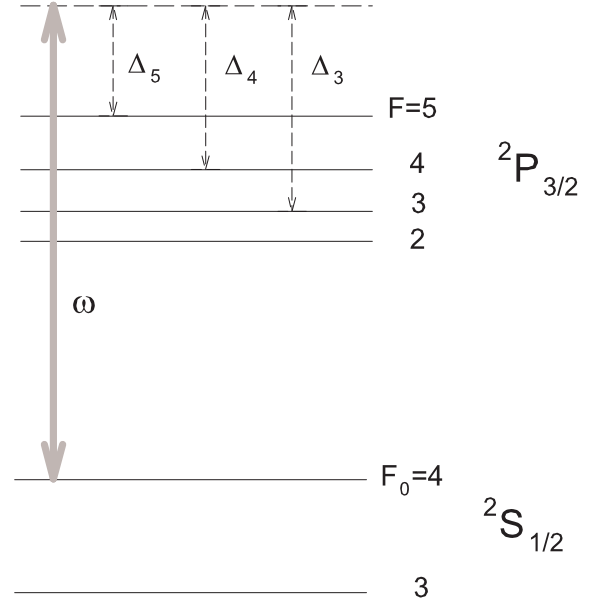}
\caption{Energy level diagram of $D_2$-line of ${}^{133}$Cs:
$\Delta_5$, $\Delta_4$, and $\Delta_3$ are the frequency detunings
of the probe light from hyperfine transitions participating in the
process.}
\label{Fig.2}%
\end{figure}

The scheme of energy levels of ${}^{133}$Cs is shown in figure
\ref{Fig.2}. Skipping technical details of calculation of
off-resonant cross section $\sigma_{\Delta}$ (on an atom in the
Zeeman sublevel $F_0=4,M_0=4$ of its ground state) and of
estimation of the elementary Faraday angle $\beta$, we present the
final result. The square variance of any output Stokes component
$\Xi_i^{out}$, where $i=1,2$, is given by
\begin{equation}
\left\langle\left[\int_0^T%
\hat{\Xi}_i^{out}(t)dt\right]^2\right\rangle\;=\;%
\bar{\Xi}_3T\,\left[1\,+\,\xi_i(J)\right]%
\label{6.18}%
\end{equation}%
where the Mandel parameter $\xi_i(J)$, considered as function of
the total collective angular momentum $J\approx \bar{J}_x$, shows
the relative deviation from the shot-noise level. This deviation
is due to the variance of the atomic state mapped onto the output
light. For $\hat{\Xi}_1^{out}$ Stokes component the original value
of the Mandel parameter $\xi_1$ transforms as
\begin{eqnarray}
\xi_1(J)&=&\xi_1\,+\,2\eta\,%
f\left(\frac{\gamma}{\Delta_5},\frac{\gamma}{\Delta_4},%
\frac{\gamma}{\Delta_3}\right)\,\beta J%
\nonumber\\%
&\equiv&\xi_1\,+\,\kappa^2
\label{6.19}%
\end{eqnarray}%
where the second line defines the dimensionless parameter $\kappa$
responsible for coupling of the field and atomic subsystems
described in terms of their canonical variable, see \cite{Polzik}.
For $\hat{\Xi}_2^{out}$ component the Mandel parameter preserves
its magnitude, so that $\xi_2(J)=\xi_2$. In (\ref{6.19}) and
throughout we approximate the input fluctuations as Markovian
delta-correlated process in the lower frequency domain
\begin{equation}
\left\langle\hat{\Xi}_i^{in}(t')\hat{\Xi}_i^{in}(t)\right\rangle\;=\;%
\bar{\Xi}_3\,(1+\xi_i)\,\delta(t'-t)%
\label{6.20}%
\end{equation}%
and assume the Poissonian (coherent state) square variance of the
angular momentum $\langle J_z^2\rangle=\bar{J}_x/2=J/2$ for
original coherent atomic spin state. Second term in (\ref{6.19})
denotes the contribution of the Faraday effect itself and function
$f(\ldots)$ denotes the product
\begin{eqnarray}
\beta\,\frac{S_0}{\sigma_{\Delta}}&\equiv&f\left(\frac{\gamma}{\Delta_5},%
\frac{\gamma}{\Delta_4},\frac{\gamma}{\Delta_3}\right)%
\nonumber\\%
&=&\frac{\frac{11}{60}\frac{\gamma}{\Delta_5}-\frac{7}{320}\frac{\gamma}{\Delta_4}%
-\frac{7}{192}\frac{\gamma}{\Delta_3}}%
{\frac{3}{10}\frac{\gamma^2}{\Delta_5^2}+\frac{7}{10}\frac{\gamma^2}{\Delta_4^2}}%
\label{6.21}%
\end{eqnarray}%
where $\gamma$ is the rate of the spontaneous decay of the upper
state and $\Delta_j=\omega-\omega_{j4}$ with $j=3,4,5$ are the
frequency detunings of the probe light from each exciting
hyperfine transition, see figure \ref{Fig.2}.

\begin{figure}[tp]
\includegraphics{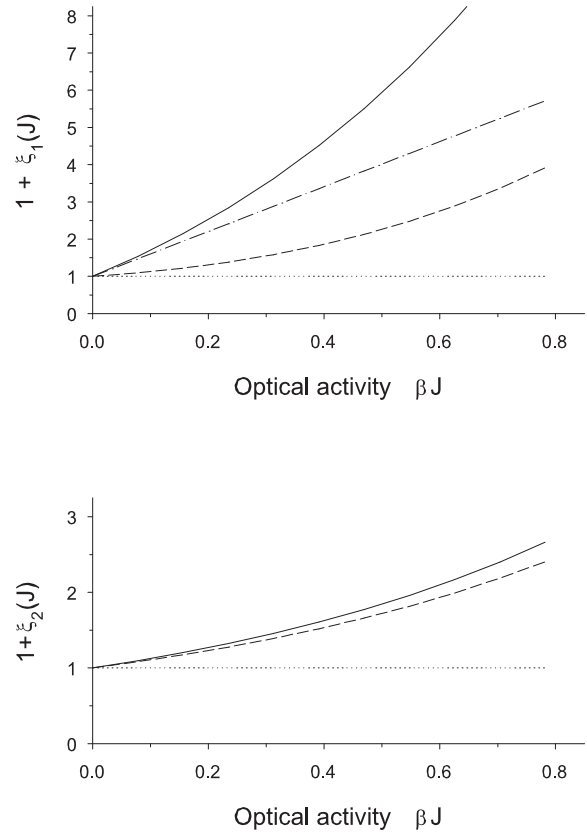}
\caption{The Mandel parameters for the output square variances of
the Stokes components $\Xi_1^{out}$ (upper plot) and $\Xi_2^{out}$
(lower plot) as a function of optical activity of the sample for
the frequency offset $\Delta_5=700\, MHz$. The dotted curve
indicates the original shot-noise level, the dashed curve is the
atomic signal coming from the first term in (\ref{6.8}), the
dashed-dotted linear dependence is the Faraday approximation
(\ref{6.19}), the solid curve is the complete output variance.}
\label{Fig.3}%
\end{figure}

\begin{figure}[tp]
\includegraphics{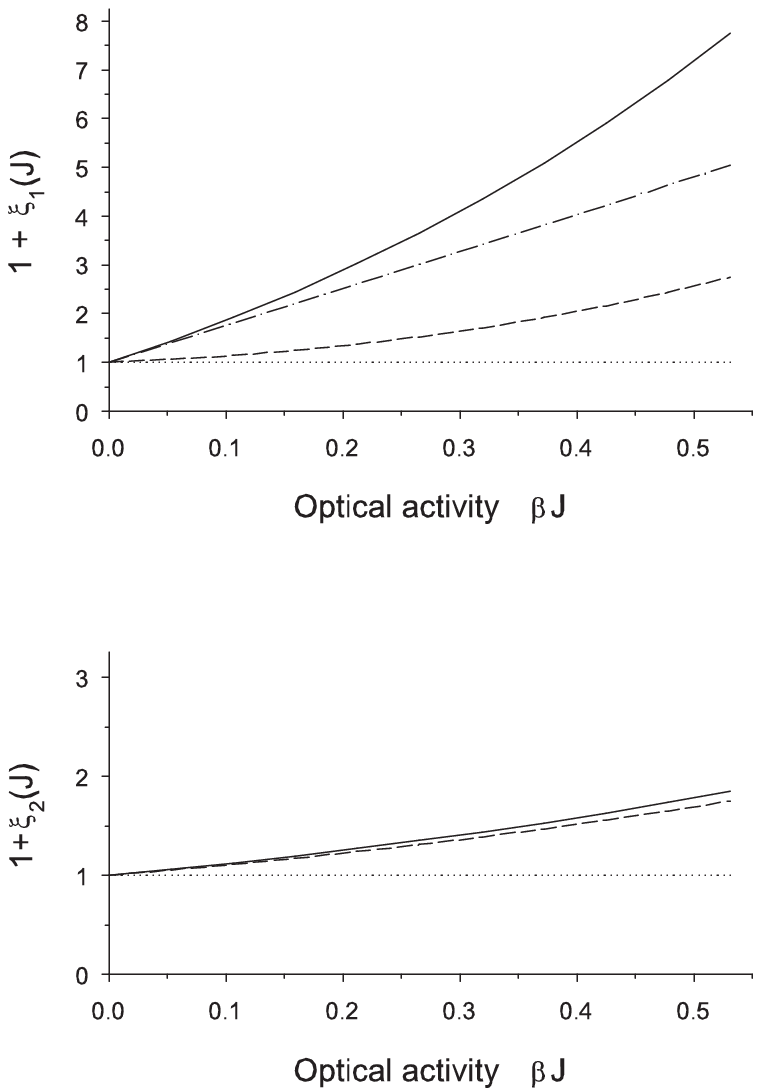}
\caption{Same as in figure \ref{Fig.3} for the frequency offset
$\Delta_5=1000\, MHz$.}
\label{Fig.4}%
\end{figure}

\begin{figure}[tp]
\includegraphics{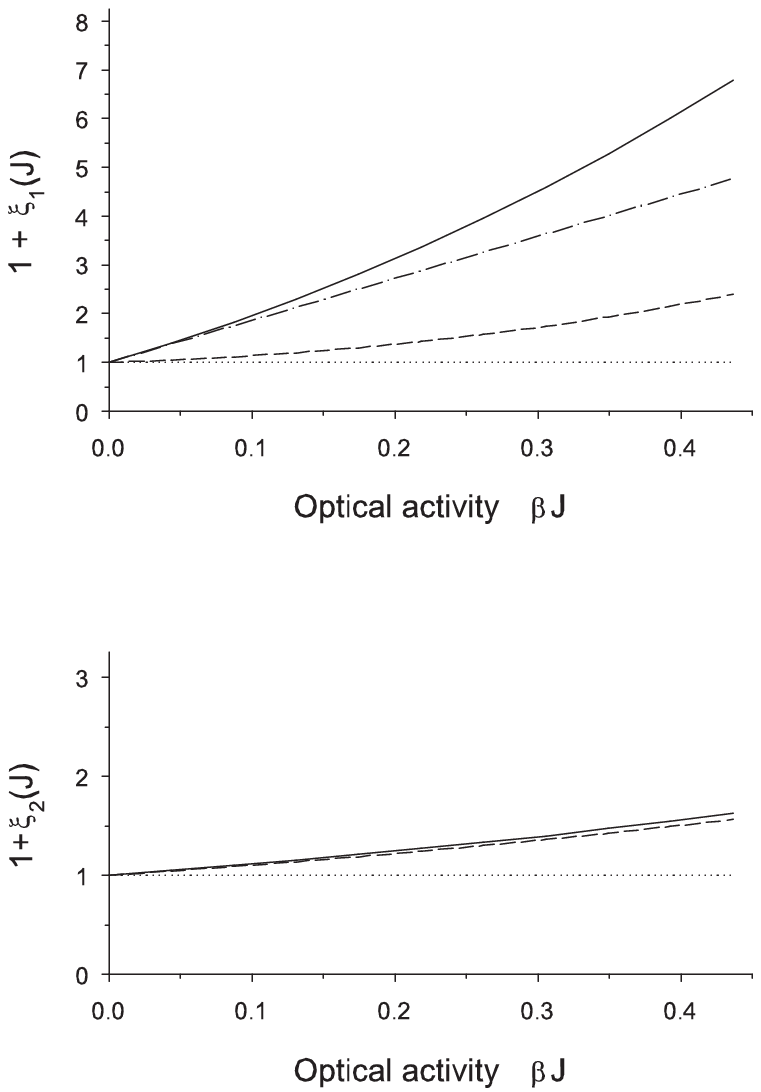}
\caption{Same as in figure \ref{Fig.3} for the frequency offset
$\Delta_5=1200\, MHz$.}
\label{Fig.5}%
\end{figure}

Note that, as a consequence of neglecting the alignment effects,
the total angular momentum (or a number of atoms) contributes in
the output Mandel parameter only in combination $\beta J$. This
parameter is the optical activity of the sample, i.e. the angle of
Faraday rotation of the planar wave for the ensemble perfectly
oriented along the probe beam, see figure \ref{Fig.1}. This
linearity (see dashed-dotted lines in figures
\ref{Fig.3}-\ref{Fig.5}) has been used as the benchmark for the
determination of the projection noise level of atoms in all the
work related to quantum state generation with atomic ensembles,
see for example \cite{Polzik}.

As follows from the above discussion of the role of spontaneous
emission, a reasonable choice of parameters for spin squeezing
corresponds to $\beta J\sim F_0\eta f(\ldots)=4\eta f(\ldots)$.
Under typical conditions for the frequency detuning $\Delta_5\sim
1000\,MHz$ and $\eta\sim 0.1$ the optical activity can be $\beta
J\sim 1$, which can provide spin squeezing far below the standard
quantum limit in (\ref{6.18}).

In order to calculate the corrections to the variances of the
output Stokes components for a realistic alkali atom at a finite
detuning the alignment associated effects have to be included in
calculations. This can be done only numerically by means of the
transformations (\ref{6.8}). In figures \ref{Fig.3}-\ref{Fig.5} we
plot the results of these calculations. The variances of the
Stokes components $\hat{\Xi}_1^{out}$ and $\hat{\Xi}_2^{out}$ are
shown as a function of the optical activity of the sample $\beta
J$, calculated for different frequency offsets $\Delta_5= 700,\,
1000,\, 1200\, MHz$ and for $\eta=0.1$. In the figures the solid
curves represent the complete result within the described model,
whereas dashed curves are computed only including the partial
contribution coming from the first term in Eq.(\ref{6.8}). These
curves show the excessive atomic spin fluctuations, mapped onto
the Stokes collective variable of the transmitted light, beyond
the level of the transformed original light fluctuations and
beyond the level of the spin (orientation) projection noise. The
incoming light is taken to be in a coherent state with Poissonian
statistics and with $\xi_1=\xi_2=0$. The enhancement of the light
noise is caused by higher orders of the interaction process.
Considered at certain value of $\beta J$ the difference between
the complete variance and the background level shows the
efficiency of the Faraday detector as a spin squeezer. The scheme
works the better the higher this difference is. However, because
of the alignment the squeezing has to be considered not for a
collective angular momentum but for a more complex type of a
spatial spin mode, which is defined by the input-output
transformations (\ref{6.8}) and (\ref{6.9}). Squeezing and
entanglement for such spatial modes will be considered elsewhere.

It is instructive to compare the result of numerical calculations
with the approximation ignoring the alignment effects. The
dashed-dotted linear dependencies in figures
\ref{Fig.3}-\ref{Fig.5} reproduces the ideal result described by
Eq.(\ref{6.19}). As we see, for small values of $\beta J$ the
Faraday approximation fits the exact solution with rather good
accuracy and the deviation of $\xi_2$ from zero is quite small.
This means that for the sample characterized by rather small
optical activity the spin one-half approximation for the
multi-level cesium atom is self-consistent and potentially good
for describing the real experimental situation. Hence for such a
sample the squeezed spin standing wave can be approximated by the
collective angular momentum. Let us point out here that in the
existing experiment \cite{JKP} the optical activity was $\beta
J<0.2$ and the alignment correction was not so important. However,
as clearly seen for large $\beta J$ the difference between the
input-output transformations in their general form and
approximation (\ref{6.14}) becomes quite important.

\section{Conclusion}
We have considered the quantum theory of coherent forward
scattering of light by an ensemble of multi-level atoms polarized
in their angular momenta. As a result of such a scattering process
the quantum states of the field and atomic subsystems are
transformed into an entangled state. In our discussion of the
process we followed the effective Hamiltonian approach, which in
the semiclassical form is normally applied for studying optical
pumping processes after adiabatic elimination of the excited
state. Compared to earlier studies of entanglement of light and
atomic ensembles, we have derived the effective Hamiltonian in a
more general form for the atoms with an arbitrary total magnetic
momentum. Towards this end, we have found the physical conditions
under which the analysis can be simplified by introducing a finite
number of collective variables for light and atoms.

We showed that under certain conditions the cooperative
atoms-field dynamics can be properly described by the wave-type
coupled equations for the space-time evolution of the collective
Heisenberg operators of the field and atomic subsystems. In these
equations an infinite set of atom-field operators is truncated via
introduction of slowly varying collective modes. The coupled
equations  for the time evolution of spatially dependent operators
of the Stokes components of light and of the macroscopic
fluctuations of the collective atomic angular momentum are written
in a closed form. In the general case the coupled dynamics of the
atom and field operators manifests itself in a spin polariton wave
created in the sample. The fluctuating components of atoms and
field become strongly entangled in the polariton wave in space and
time. Such spin polariton waves initiated by radiative forces are
different from the collective spin dynamics existing in the
spin-polarized quantum gas \cite{BLL}. They are also different
from the polariton modes discussed in \cite{Lukin} where
polaritons of the combined atom-light state are introduced. In our
case the quantum entanglement arises from the interaction between
the internal collective polarization degrees of freedom of light
and atomic subsystems. In particular our analysis yields the
input-output transformations for the Heisenberg operators of the
collective variables of light and atomic spins after the whole
interaction cycle. Our results suggest that a successful approach
of using off-resonant light-atomic ensemble interaction for
quantum information processing can become even more fruitful with
the use of a multi-mode type of entanglement provided by spin
polariton atomic variables.

Numerical simulations demonstrate the importance of the developed
formalism for application to a realistic experimental situation.
We considered a well known example, when the Faraday rotation is
used as a non-demolishing measurement of transverse fluctuations
of the collective atomic spin and can be utilized as a physical
mechanism for the spin squeezing in an ensemble of spin polarized
atoms. We tested the validity of the spin one-half approximation
for a realistic Cesium atom, which is normally used to describe
the interaction with a far off-resonant probe light around $D_1$
or $D_2$ transitions of alkali atoms. As shown, for small values
of optical activity this approximation is self-consistent and
deviates negligibly from the calculations based on a general
solution. However for the samples with high optical activity,
where quantum correlations become strong, there is an important
quantitative as well as qualitative difference between the general
solution and the model of spin one-half atoms.

\section*{Acknowledgments}

Financial support for this work was provided by INTAS (Grant INFO
00-479), by Danish National Research Foundation, by the EU grant
COVAQIAL, and by the North Atlantic Treaty Organization
(PST-CLG-978468). D.V.K. would like to acknowledge financial
support from the Delzell Foundation, Inc.

\appendix
\section{Transformation of the electric field in a single
scattering} \label{s2.1}

The unperturbed electric field operator in the origin of the
coordinate frame coupled with a scattering atom, which drifts with
velocity $\mathbf{v}$, is given by
\begin{eqnarray}
\hat{\mathbf{E}}_{0}(t) &=&\sum\limits_{\mathbf{k}\mu }\,%
\left(\frac{2\pi\hbar \omega _{k}}{\mathcal{V}}\right) ^{1/2}\,%
(-i)\,\left[ \mathbf{e}_{\mathbf{k}\mu }^{\ast }\,%
a_{\mathbf{k}\mu }^{\dagger }\,e^{i(\omega _{k}-%
\mathbf{kv})t}\right.%
\nonumber \\%
&&\left.-\;\mathbf{e}_{\mathbf{k}\mu }\,%
a_{\mathbf{k}\mu}\,e^{-i(\omega _{k}-\mathbf{kv})t}\right]%
\nonumber \\%
&=&\hat{\mathbf{E}}_{0}^{(-)}(t)\;+\;\hat{\mathbf{E}}_{0}^{(+)}(t)%
\label{2.1}%
\end{eqnarray}%
where $a_{\mathbf{k}\mu }^{\dagger }$, $a_{\mathbf{k}\mu }$ are
respectively the operators of creation and annihilation of the
photon with wavevector $\mathbf{k}$ and in the polarization state
$\mathbf{e}_{\mathbf{k}\mu }$. $\mathcal{V}$ is the quantization
volume. The second line in (\ref{2.1}) defines negative and
positive frequency components of the electric field.

The dipole-type interaction operator of the atom with the electric
field is given by
\begin{equation}
\hat{V}(t)=-\hat{\mathbf{d}}(t)\,\hat{\mathbf{E}}_{0}(t)%
\label{2.2}%
\end{equation}%
where the operator for the atomic dipole moment
$\hat{\mathbf{d}}(t)$ and for the electric field are defined in
the interaction representation. Based on a perturbation theoretic
approach, the exact solution for the Heisenberg operator can be
written as the following expansion
\begin{equation}
\hat{\mathbf{E}}(t)=\hat{\mathbf{E}}_{0}(t)\;+\;%
\hat{\mathbf{E}}_{1}(t)\;+\;%
\hat{\mathbf{E}}_{2}(t)\;+\;\ldots%
\label{2.3}%
\end{equation}%
which can be also written for positive $\hat{\mathbf{E}}^{(+)}(t)$
and negative $\hat{\mathbf{E}}^{(-)}(t)$ frequency components. We
assume that the wave function of the joint atom-field system
describes the combined state where an atom occupies the ground
state and the electric field is in a weak quasi-coherent state
non-saturating the atomic transition. Then the correction of the
first order in (\ref{2.3}) will disappear after averaging over the
wave function. Thus the second order correction gives us the main
contribution, since it is responsible for the scattering process.
The second order term in (\ref{2.3}) can be written as follows
\begin{eqnarray}
\hat{\mathbf{E}}_{2}(t)&=&-\frac{1}{\hbar^{2}}\,%
\int_{0}^{t}dt^{\prime \prime}%
\int_{t^{\prime \prime }}^{t}dt^{\prime }\,%
\nonumber\\%
&&\times\!\left[ \hat{\mathbf{d}}(t^{\prime \prime })\,%
\hat{\mathbf{E}}_{0}(t^{\prime \prime }),\,%
\left[ \hat{\mathbf{d}}(t^{\prime })\,%
\hat{\mathbf{E}}_{0}(t^{\prime }),\,%
\hat{\mathbf{E}}_{0}(t)\right] \right]%
\label{2.4}%
\end{eqnarray}%
As follows from this expression, in a complete dynamical
description of the process there is a memory of initial conditions
in the formal expansion of perturbation theory. However, for
non-saturating fields this solution can be spread out over the
time $t\gg \gamma ^{-1}$, where $\gamma $ is the natural radiative
relaxation rate of the upper state. But in this case it is
necessary to take into account all the radiative correction for
the retarded and advanced Green functions of the decaying upper
atomic state. This can be done by introducing the natural decay
law into the time behavior of these functions. Then the integral
(\ref{2.4}) loses its dependence on the lower limit as on initial
time coordinate. Let us also point out that there is only
conventional choice of initial time $t_0\to 0$ to coordinate the
Heisenberg and Schr\"{o}dinger representations. The real physical
conditions can be arranged as the wave front of probe radiation
(expressed by expectation values of any products of freely
Heisenberg operators of the field) could arrive to the interaction
area at any time after $t_0=0$. Thus we can always think that
observation time $t$ satisfies inequality $t\gg \gamma^{-1}$. At
the same time it is important to recognize that by including the
radiation decay into the Green functions of the upper atomic
states we average their time evolution and ignore any random fast
variation of the field and atomic operators associated with high
frequencies of the field continuum.

The integral (\ref{2.4}) gives us the solution for the electric
field in the origin of the frame coupled with a moving atom. But
in the zero order of relativistic effects, when only retardation
effects in the radiation zone have to be taken into consideration,
this solution coincides with the electric field in the laboratory
frame at the point of atom location as well as in the small
vicinity of this point. Then we can obtain the solution for any
point in the laboratory frame by using the propagation law in free
space. By this procedure one obtains the following expansion for
the positive frequency component of the electric field operator in
the radiation zone ($r\gg \lambdabar$) of the scattering atom
\begin{equation}
\hat{\mathbf{E}}^{(+)}(\mathbf{r},t)=%
\hat{\mathbf{E}}_{0}^{(+)}(\mathbf{r},t)\,+\,%
\hat{\mathbf{E}}_{2}^{(+)}(\mathbf{r},t)\,+\,\ldots%
\label{2.5}%
\end{equation}%
where
\begin{widetext}%
\begin{equation}
\hat{\mathbf{E}}_{0}^{(+)}(\mathbf{r},t)=%
\sum\limits_{\mathbf{k}\mu}\,%
\left( \frac{2\pi \hbar \omega _{k}}{\mathcal{V}}\right)^{1/2}\,%
e^{-i(\omega _{k}t-\mathbf{k}\mathbf{r})}\,%
i\,\mathbf{e}_{\mathbf{k}\mu }\,a_{\mathbf{k}\mu }\,%
\label{2.6}
\end{equation}%
and
\begin{eqnarray}
\hat{\mathbf{E}}_{2}^{(+)}(\mathbf{r},t)&=&%
\sum\limits_{m,m^{\prime}}\sum\limits_{n}%
\sum\limits_{\mathbf{k}\mu }\,%
\left( \frac{2\pi\hbar \omega _{k}}{\mathcal{V}}\right)^{1/2}\,\frac{1}{ir}\,%
e^{-i\omega ^{\prime }t+ik^{\prime}r}\,%
|m^{\prime }\rangle \langle m|\,a_{\mathbf{k}\mu }%
\nonumber\\%
&&\times \frac{{\omega ^{\prime }}^{2}}{\hbar c^{2}}%
\left[ -\frac{i(\mathbf{d}_{\perp })_{nm}\,%
(\mathbf{d}\mathbf{e}_{\mathbf{k}\mu })_{m^{\prime }n}}%
{i(\omega ^{\prime }+\omega _{nm}-\mathbf{k^{\prime }v})}\;+\;%
\frac{i(\mathbf{d}_{\perp })_{m^{\prime }n}\,%
(\mathbf{d}\mathbf{e}_{\mathbf{k}\mu })_{nm}}%
{i(\omega _{k}-\omega _{nm}-\mathbf{k}\mathbf{v})-\gamma_{n}/2}\right]%
\label{2.7}%
\end{eqnarray}%
\end{widetext}%
and the negative frequency component is given by the Hermitian
conjugation: $\mathbf{E}^{(-)}(\mathbf{r},t)=
\mathbf{E}^{(+)\dagger}(\mathbf{r},t)$. Here the origin of the
frame is chosen in the location of the atom, which is assumed to
be unchanged during the light propagation time $r/c$. The
scattered light frequency $\omega ^{\prime }$ is defined here via
the input frequency $\omega _{k}$, and the Raman shift $\omega
_{m^{\prime }m}$ for atomic transition $|m\rangle \rightarrow
|m^{\prime }\rangle $ There is an additional Doppler shift caused
by atomic motion, given by
\begin{equation}
\omega^{\prime }\equiv \omega _{k^{\prime }}=%
\omega_{k}\,-\,\omega _{m^{\prime }m}\,+\,%
(\mathbf{k}^{\prime}-\mathbf{k})\mathbf{v}%
\label{2.8}%
\end{equation}%
The wave vector $\mathbf{k}^{\prime }=\omega ^{\prime
}\mathbf{r}/cr$, but in the right hand side of (\ref{2.8}) the
Doppler correction is neglected and it is assumed that
$\mathbf{k}^{\prime }\approx \omega \mathbf{r}/cr$. The transition
dipole moments in (\ref{2.7}) are defined in the Schr\"{o}dinger
representation and its transverse component is given by
\begin{equation}
\hat{\mathbf{d}}_{\perp }=\hat{\mathbf{d}}\,-\,%
(\hat{\mathbf{d}}\,\mathbf{k}^{\prime })%
\frac{\mathbf{k}^{\prime }}{{k^{\prime }}^{2}}%
\label{2.9}%
\end{equation}%
The sum over $n$ is expanded over all possible excited transitions
characterized by natural linewidths $\gamma _{n}$, but as a
practical matter, the sum can be restricted to the most
significant resonance transitions and\ the frequency $\omega
_{k}\equiv \omega $ can be associated with the frequency of the
incident mode.

In spite both the terms in (\ref{2.6}) are well known in the
scattering theory, see \cite{BLP}, we claim that only the latter
term, constituted with rotating wave approximation, should be left
in aiming the following generalization to multiple scattering
process. Indeed, the fist term contributed in (\ref{2.6}) reveals
the quantum mechanical phenomenon than the scattered photon is
created in advance than incoming photon was annihilated. This
process has very small, and in fact negligible, amplitude and
should be ignored if one stays with semiclassical-type
understanding of multiple scattering process as sequence of
successive scattering events. Moreover, as shown in \cite{KSKSH},
the electric field transformation in the presence of one scatterer
in form (\ref{2.5})-(\ref{2.7}) preserves commutation relation, i.
e. being unitary transformation only in rotating wave
approximation.

In the rotating wave approximation the perturbation theory
solution (\ref{2.5})-(\ref{2.7}) satisfies the following equation
\begin{equation}
\triangle \hat{\mathbf{E}}^{(+)}(\mathbf{r},t)\,-\,%
\frac{1}{c^2}\frac{\partial^2}{\partial t^2}%
\hat{\mathbf{E}}^{(+)}(\mathbf{r},t)\;=\;%
\frac{4\pi}{c^2}\frac{\partial^2}{\partial t^2}%
\hat{\mathbf{P}}^{(+)}_{\bot}(\mathbf{r},t)%
\label{2.10}%
\end{equation}%
where
\begin{equation}
\hat{\mathbf{P}}^{(+)}_{\bot}(\mathbf{r},t)\;=\;%
\int\frac{d^3k'}{(2\pi)^3}\;e^{i\mathbf{k}'\mathbf{r}}\,%
\hat{\mathbf{P}}^{(+)}_{\bot}(\mathbf{k}',t)%
\label{2.11}%
\end{equation}%
and in turn
\begin{eqnarray}
\hat{\mathbf{P}}^{(+)}_{\bot}(\mathbf{k}',t)&=&%
\sum\limits_{m,m^{\prime}}\sum\limits_{n}%
\sum\limits_{\mathbf{k}\mu }\,%
\left( \frac{2\pi\hbar \omega _{k}}{\mathcal{V}}\right)^{1/2}%
\phantom{e^{-i\omega ^{\prime }t}\,i\,a_{\mathbf{k}\mu }}%
\nonumber\\%
&&\times\,e^{-i\omega ^{\prime }t}\,%
|m^{\prime }\rangle \langle m|\,i\,a_{\mathbf{k}\mu }%
\nonumber\\%
&&\frac{(\mathbf{d}_{\perp })_{m^{\prime }n}\,%
(\mathbf{d}\mathbf{e}_{\mathbf{k}\mu })_{nm}}%
{-\hbar(\omega _{k}-\omega _{nm}-\mathbf{k}\mathbf{v})\,-\,%
i\hbar\gamma_{n}/2}%
\label{2.12}%
\end{eqnarray}
is the operator of transverse component of atomic polarization
responding on external field. Frequency $\omega'$ is given by
Eq.(\ref{2.8}), but wavevector $\mathbf{k}'$ is independent
variable here.

In above discussion the interaction time $t$ was assumed to be
short enough for validity of the perturbation theory approach.
Therefore in the derived equations the dyadic-type operators
$|m^{\prime }\rangle \langle m|$ of the low atomic state can be
selected as referred to interaction representation
\begin{equation}
|m^{\prime }\rangle \langle m|_0\,(t)\;=\;%
e^{i\omega_{m'm}t}\,|m^{\prime }\rangle \langle m|%
\label{2.13}%
\end{equation}%
But this operators only slightly differs from their Heisenberg
analogs during the short interaction time. Therefore the evolution
of the electric field operator can be extended up to arbitrary
time $t$, if the final equations of this section are modified as
follows. All the dyadic-type operators should be changed by the
corresponding Heisenberg operators with keeping the complete
dynamic evolution up to moment $t$
\begin{equation}
|m^{\prime }\rangle \langle m|_0(t)\;\to\;%
|m^{\prime }\rangle \langle m|(t)%
\label{2.14}
\end{equation}
The drift of the atom in space cannot be further ignored and
instead of origin of the laboratory coordinate frame we should
assume its actual location associated with its classical motion
$\mathbf{r}_a=\mathbf{r}_a(t)$. Equation (\ref{2.10}) stays valid
and unchanged but atomic polarization (\ref{2.12}) modifies to
\begin{eqnarray}
\hat{\mathbf{P}}^{(+)}_{\bot}(\mathbf{k}',t)&=&%
\sum\limits_{m,m^{\prime}}\sum\limits_{n}%
\sum\limits_{\mathbf{k}\mu }\,%
\left( \frac{2\pi\hbar \omega _{k}}{\mathcal{V}}\right)^{1/2}%
\phantom{e^{i(\mathbf{k} - \mathbf{k}')\mathbf{r}_a(t)}\,}%
\nonumber\\%
&&\times\,e^{i(\mathbf{k} - \mathbf{k}')\mathbf{r}_a(t)}\,%
|m^{\prime }\rangle \langle m|(t)\,i\,a_{\mathbf{k}\mu }(t)%
\nonumber\\%
&&\frac{(\mathbf{d}_{\perp })_{m^{\prime }n}\,%
(\mathbf{d}\mathbf{e}_{\mathbf{k}\mu })_{nm}}%
{-\hbar(\omega _{k}-\omega _{nm}-\mathbf{k}\mathbf{v}_a)\,-\,%
i\hbar\gamma_{n}/2}%
\label{2.15}%
\end{eqnarray}
and $\mathbf{v}_a=\dot{\mathbf{r}}_a(t)$. In substituting
(\ref{2.15}) into Eq.(\ref{2.10}) the time derivation should be
done only for the fast oscillating components of the Heisenberg
operators and for the exponential factors.

\section{Linearized dynamics of atomic angular momenta}\label{a}

The operators of irreducible components can be expressed in terms
of the operators of atomic angular momenta. For any $a$-th atom of
ensemble the orientation vector is given by
\begin{equation}
\hat{T}_{1Q}^{(a)}\;=\;\frac{\sqrt{3}}%
{[j_0(j_0+1)(2j_0+1)]^{1/2}}\,\hat{j}_Q^{(a)}%
\label{a.1}%
\end{equation}%
and the alignment tensor is given by
\begin{eqnarray}
\hat{T}_{2Q}^{(a)}&=&\frac{\sqrt{15}}%
{[{2j_0(j_0+1)(2j_0-1)(2j_0+1)(2j_0+3)}]^{1/2}}\,%
\nonumber\\%
&&\times\sum_{qq'}\,C^{2Q}_{1q\,1q'}\,%
\left[\hat{j}_q^{(a)}\hat{j}_{q'}^{(a)}\,+\,%
\hat{j}_{q'}^{(a)}\hat{j}_q^{(a)}\right.%
\nonumber\\%
&&\left.\phantom{C^{2Q}_{1q\,1q'}}-\;%
(-)^q\delta_{q,-q'}\frac{2}{3}j_0(j_0+1)\right]%
\label{a.2}%
\end{eqnarray}
where $\hat{j}_q^{(a)}$ are the cyclic components of the operator
vector of angular momentum, which are defined by their Cartesian
components as follows
\begin{eqnarray}
\hat{j}_0^{(a)}&=&\hat{j}_z^{(a)}%
\nonumber\\%
\hat{j}_{\pm 1}^{(a)}&=&\mp\frac{1}{\sqrt{2}}%
\left[\hat{j}_x^{(a)}\,\pm\,i\hat{j}_y^{(a)}\right]%
\label{a.3}%
\end{eqnarray}%
see \cite{VMK}.

By substituting subsequently (\ref{a.1})-(\ref{a.3}) into
equations (\ref{5.7}) the latter can be straightforwardly
transformed to the set of nonlinear equations containing only the
operators of atomic angular momenta
\begin{widetext}%
\begin{eqnarray}
\dot{\hat{j}}{}_{z}^{(a)}(t)&=&%
\Omega_0\,\hat{j}_{y}^{(a)}(t)\;-\;%
\bar{\alpha}_2\,(-)^{2j_0}\frac{5}{[(2j_0-1)(2j_0+3)]^{1/2}}\,%
\left\{\begin{array}{ccc}%
2 & 1 & 2\\ j_0 & j_0 & j_0%
\end{array}\right\}\,\hat{\Xi}_3(z_a(t),t)\,%
\left[\hat{j}_{x}^{(a)}(t)\hat{j}_{y}^{(a)}(t)\,+\,%
\hat{j}_{y}^{(a)}(t)\hat{j}_{x}^{(a)}(t)\right]%
\nonumber\\%
&&+\;\bar{\alpha}_2\,(-)^{2j_0}\frac{5}{[(2j_0-1)(2j_0+3)]^{1/2}}\,%
\left\{\begin{array}{ccc}%
2 & 1 & 2\\ j_0 & j_0 & j_0%
\end{array}\right\}\,%
\hat{\Xi}_1(z_a(t),t)\,%
\left[\hat{j}_{x}^{(a)2}(t)\,-\,%
\hat{j}_{y}^{(a)2}(t)\right],%
\nonumber\\%
\nonumber\\%
\dot{\hat{j}}{}_{y}^{(a)}(t)&=&%
-\Omega_0\,\hat{j}_{z}^{(a)}(t)\;+\;%
\bar{\alpha}_2\,(-)^{2j_0}\frac{5}{2[(2j_0-1)(2j_0+3)]^{1/2}}\,%
\left\{\begin{array}{ccc}%
2 & 1 & 2\\ j_0 & j_0 & j_0%
\end{array}\right\}%
\nonumber\\%
&&\times\left[\hat{\Xi}_0(z_a(t),t)\,+\,\hat{\Xi}_3(z_a(t),t)\right]%
\left[\hat{j}_{x}^{(a)}(t)\hat{j}_{z}^{(a)}(t)\,+\,%
\hat{j}_{z}^{(a)}(t)\hat{j}_{x}^{(a)}(t)\right]%
\nonumber\\%
&&+\;\bar{\alpha}_2\,(-)^{2j_0}\frac{5}{2[(2j_0-1)(2j_0+3)]^{1/2}}\,%
\left\{\begin{array}{ccc}%
2 & 1 & 2\\ j_0 & j_0 & j_0%
\end{array}\right\}\,\hat{\Xi}_1(z_a(t),t)\,%
\left[\hat{j}_{y}^{(a)}(t)\,\hat{j}_{z}^{(a)}(t)\,+\,%
\hat{j}_{z}^{(a)}(t)\,\hat{j}_{y}^{(a)}(t)\,\right]%
\nonumber\\%
&&+\;\bar{\alpha}_1\,\frac{\sqrt{3}}{2[j_0(j_0+1)(2j_0+1)]^{1/2}}\;%
\hat{\Xi}_2(z_a(t),t)\,\hat{j}_{x}^{(a)}(t),%
\nonumber\\%
\nonumber\\%
\dot{\hat{j}}{}_{x}^{(a)}(t)&=&%
-\;\bar{\alpha}_2\,(-)^{2j_0}\frac{5}{2[(2j_0-1)(2j_0+3)]^{1/2}}\,%
\left\{\begin{array}{ccc}%
2 & 1 & 2\\ j_0 & j_0 & j_0%
\end{array}\right\}%
\nonumber\\%
&&\times\left[\hat{\Xi}_0(z_a(t),t)\,-\,\hat{\Xi}_3(z_a(t),t)\right]%
\left[\hat{j}_{y}^{(a)}(t)\hat{j}_{z}^{(a)}(t)\,+\,%
\hat{j}_{z}^{(a)}(t)\hat{j}_{y}^{(a)}(t)\right]%
\nonumber\\%
&&-\;\bar{\alpha}_2\,(-)^{2j_0}\frac{5}{2[(2j_0-1)(2j_0+3)]^{1/2}}\,%
\left\{\begin{array}{ccc}%
2 & 1 & 2\\ j_0 & j_0 & j_0%
\end{array}\right\}\,\hat{\Xi}_1(z_a(t),t)\,%
\left[\hat{j}_{x}^{(a)}(t)\,\hat{j}_{z}^{(a)}(t)\,+\,%
\hat{j}_{z}^{(a)}(t)\,\hat{j}_{x}^{(a)}(t)\,\right]%
\nonumber\\%
&&-\;\bar{\alpha}_1\,\frac{\sqrt{3}}{2[j_0(j_0+1)(2j_0+1)]^{1/2}}\;%
\hat{\Xi}_2(z_a(t),t)\,\hat{j}_{y}^{(a)}(t)%
\label{a.4}%
\end{eqnarray}%
\end{widetext}%
In their general form these equations are quite complicated and
not closed because of their nonlinear structure, but they can be
simplified and linearized in the following assumptions. The
dynamics of operators $\hat{j}_{x}^{(a)}(t)$ is driven only by
those terms, which, being averaged, have a quadratic scale over
fluctuations of the field and atomic variables. Recall that
$\bar{\Xi}_0=\bar{\Xi_3}$. Physically this means that there is no
coherent process demolishing the original spin orientation of
single atom along $x$ direction. So far we neglected any
possibilities of incoherent scattering our analysis has to be
restricted by assumption that in average
$\bar{j}_{x}^{(a)}(t)=j_0$. Moreover, since this observable has a
maximal possible expectation value and it has no deviation from
$j_0$ in the lower orders of weak external perturbations, it is
allowed to approximate the Heisenberg operator
$\hat{j}_{x}^{(a)}(t)$ when it appears in linear or in the squared
nonlinear form by its non-perturbed projector onto atomic
wavefunction
\begin{equation}
\hat{j}_{x}^{(a)}(t)\;\to\;%
j_0|j_0,j_0\rangle\langle j_0,j_0|(t)\sim {\rm const}_t%
\label{a.5}%
\end{equation}%
But while substituting it in the operators' products one has to
follow the rule
\begin{eqnarray}
\hat{j}_{x}^{(a)}(t)\hat{j}_{y}^{(a)}(t)\,+\,%
\hat{j}_{y}^{(a)}(t)\hat{j}_{x}^{(a)}(t)&\to&%
(2j_0-1)\,\hat{j}_{y}^{(a)}(t)%
\phantom{(A.6)}%
\nonumber\\%
\hat{j}_{x}^{(a)}(t)\hat{j}_{z}^{(a)}(t)\,+\,%
\hat{j}_{z}^{(a)}(t)\hat{j}_{x}^{(a)}(t)&\to&%
(2j_0-1)\,\hat{j}_{z}^{(a)}(t)%
\label{a.6}%
\end{eqnarray}
This is crucial point of linearizing procedure, it is expected
that the left and right-hand side operators are coordinated if the
spin subsystem is slightly disturbed and the time dynamics of
off-diagonal projectors $|j_0,j_0\rangle\langle j_0,j_0-1|(t)$ and
$|j_0,j_0-1\rangle\langle j_0,j_0|(t)$ is only taken into
consideration. Then orientation dynamics, described by
(\ref{a.4}), can be approximated by the following set of
linearized equations
\begin{eqnarray}
\dot{\hat{j}}_z^{(a)}(t)&\approx&(\Omega_0\,+\,\Omega_2)\,%
\hat{j}_y^{(a)}(t)\;-\;%
\bar{T}_{xy}^{(a)}\,\bar{\alpha}_2\,%
\hat{\Xi}_1(z_a(t),t)%
\nonumber\\%
\dot{\hat{j}}_y^{(a)}(t)&\approx&-(\Omega_0\,+\,\Omega_2)\,%
\hat{j}_z^{(a)}(t)\;+\;%
\frac{1}{2}\bar{T}_{x}^{(a)}\,\bar{\alpha}_1\,%
\hat{\Xi}_2(z_a(t),t)%
\nonumber\\%
\dot{\hat{j}}_x^{(a)}(t)&\approx&0
\label{a.7}%
\end{eqnarray}%
where
\begin{equation}
\Omega_2\;=\;\frac{[15(2j_0-1)]^{1/2}}%
{[2j_0(j_0+1)(2j_0+1)(2j_0+3)]^{1/2}}\;%
\bar{\alpha}_2\,\bar{\Xi}_0%
\label{a.8}%
\end{equation}%
is the light induced shift between $m=j_0$ and $m=j_0-1$
sublevels. Average value of alignment component
$\bar{T}_{xy}^{(a)}$, is defined in the body of the paper by
Eq.(\ref{5.5}). Similarly we introduced here the orientation
component $\bar{T}_{x}^{(a)}$, associated with average orientation
of $a$-th atom, as
\begin{equation}
\bar{T}_{x}^{(a)}\;=\;\frac{[3j_0]^{1/2}}%
{[(j_0+1)(2j_0+1)]^{1/2}}\;\equiv\;\bar{T}_{x}%
\label{a.9}%
\end{equation}%
Here $\bar{T}_{x}^{(a)}$ is the average orientation component of
$10$-type defined in the frame with $Z$-axis along magnetic field,
which coincides with $x$-axis in our case. This component is the
same for all the atoms of ensemble. Making the sum over all
partial equations (\ref{a.7}) we come to equations (\ref{5.9})
written for the collective vector of atomic angular momentum.

\section{Laplace transform of the atoms-field dynamical
equations}\label{b}

Let us define the Laplace images of the space-time dependent
Stokes components of the probe light and of the collective angular
momentum of atoms
\begin{eqnarray}
\hat{\Xi}_i(p,s)&=&\int_0^\infty\int_0^\infty dz\,dt%
e^{-pz-st}\,\hat{\Xi}_i(z,t)\ \ i=1,2 \phantom{(B.1)}%
\nonumber\\%
\hat{{\cal J}}_\mu(p,s)&=&\int_0^\infty\int_0^\infty dz\,dt%
e^{-pz-st}\,\hat{{\cal J}}_\mu(z,t)\ \ \mu=z,y%
\label{b.1}%
\end{eqnarray}%
where parameters $p,s>0$. Then the original system of differential
equations (\ref{6.7}) with initial and boundary conditions
(\ref{5.13}) can be transformed to the following system of linear
algebraic equations for the Laplace images
\begin{eqnarray}
\phantom{+}p\,\hat{\Xi}_1(p,s)\,+\,\kappa_2\,\hat{\Xi}_2(p,s)\,-\,%
2\beta\,\bar{\Xi}_3\,\hat{{\cal J}}_z(p,s)&=&%
\hat{\Xi}_1^{in}(s)%
\nonumber\\%
-\kappa_2\,\hat{\Xi}_1(p,s)\,+\,p\,\hat{\Xi}_2(p,s)\,+\,%
2\epsilon\,\bar{\Xi}_3\,\hat{{\cal J}}_y(p,s)&=&%
\hat{\Xi}_2^{in}(s)%
\nonumber\\%
\phantom{+}\theta_y\,\bar{{\cal J}}_x\,\hat{\Xi}_1(p,s)\,+\,%
s\,\hat{{\cal J}}_z(p,s)\,-\,%
\Omega\,\,\hat{{\cal J}}_y(p,s)&=&%
\hat{{\cal J}}_z^{in}(p)%
\nonumber\\%
-\theta_z\,\bar{{\cal J}}_x\,\hat{\Xi}_2(p,s)\,+\,%
\Omega\,\hat{{\cal J}}_z(p,s)\,+\,%
s\,\,\hat{{\cal J}}_y(p,s)&=&%
\hat{{\cal J}}_y^{in}(p)%
\nonumber\\%
&&\label{b.2}%
\end{eqnarray}%
The determinant of this system is given by
\begin{eqnarray}
\Delta&=&\Delta(p,s)\;=\;(s^2+\Omega^2)(p^2+\kappa_2^2)%
\nonumber\\%
&&+\;2(\epsilon\theta_z+\beta\theta_y)\bar{{\cal J}}_x\,\bar{\Xi}_3\,s\,p%
\,-\,2(\epsilon\theta_y+\beta\theta_z)%
\bar{{\cal J}}_x\,\bar{\Xi}_3\,\Omega\,\kappa_2%
\nonumber\\%
&&+\;4\beta\epsilon\theta_z\theta_y(\bar{{\cal J}}_x\,\bar{\Xi}_3)^2%
\label{b.3}%
\end{eqnarray}%
and its solution can be easy found by means of the Kramer's rules.

The solution is expressed as linear transformations of the Laplace
images of the boundary Stokes operators for the field and of the
initial angular momentum operators for the atoms
\begin{eqnarray}
\hat{\Xi}_i(p,s)&=&%
\sum_{j=1,2}\,%
{\cal M}_{ij}(p,s)\,\hat{\Xi}_j^{in}(s)%
\nonumber\\%
&&+\;\sum_{\nu=z,y}\,%
{\cal F}_{i\nu}(p,s)\,\hat{{\cal J}}_{\nu}^{in}(p)%
\nonumber\\%
\hat{{\cal J}}_{\mu}(p,s)&=&%
\sum_{j=1,2}\,%
{\cal G}_{\mu j}(p,s)\,\hat{\Xi}_j^{in}(s)%
\nonumber\\%
&&+\;\sum_{\nu=z,y}\,%
{\cal N}_{\mu\nu}(p,s)\,\hat{{\cal J}}_{\nu}^{in}(p)%
\label{b.4}%
\end{eqnarray}%
These transformations perform the Laplace images of the integral
transforms (\ref{6.8}) and (\ref{6.9}) introduced in the body of
the paper.

The self-transformation matrix ${\cal M}(p,s)$ of the Stokes
operators  is given by
\begin{widetext}%
\begin{equation}
{\cal M}(p,s)=\frac{1}{\Delta(p,s)}%
\left(\begin{array}{cc}%
p(s^2+\Omega^2)\,+\,2\epsilon\theta_z\,%
\bar{{\cal J}}_x\,\bar{\Xi}_3\,s, &%
-\kappa_2 (s^2+\Omega^2)\,+\,2\beta\theta_z\,%
\bar{{\cal J}}_x\,\bar{\Xi}_3\,\Omega\\%
\kappa_2(s^2+\Omega^2)\,-\,2\epsilon\theta_y\,%
\bar{{\cal J}}_x\,\bar{\Xi}_3\,\Omega, &%
p (s^2+\Omega^2)\,+\,2\beta\theta_y\,%
\bar{{\cal J}}_x\,\bar{\Xi}_3\,s%
\end{array}\right)
\label{b.5}%
\end{equation}%
\end{widetext}%
The self-transformation matrix ${\cal N}(p,s)$ of the angular
momentum operators is given by
\begin{widetext}%
\begin{equation}
{\cal N}(p,s)=\frac{1}{\Delta(p,s)}%
\left(\begin{array}{cc}%
s(p^2+\kappa_2^2)\,+\,2\epsilon\theta_z\,%
\bar{{\cal J}}_x\,\bar{\Xi}_3\,p, &%
\Omega (p^2+\kappa_2^2)\,-\,2\epsilon\theta_y\,%
\bar{{\cal J}}_x\,\bar{\Xi}_3\,\kappa_2\\%
-\Omega(p^2+\kappa_2^2)\,+\,2\beta\theta_z\,%
\bar{{\cal J}}_x\,\bar{\Xi}_3\,\kappa_2, &%
s (p^2+\kappa_2^2)\,+\,2\beta\theta_y\,%
\bar{{\cal J}}_x\,\bar{\Xi}_3\,p%
\end{array}\right)
\label{b.6}%
\end{equation}%
\end{widetext}%
The cross-transformation matrix ${\cal F}(p,s)$, which is
responsible for mapping the initial angular momentum fluctuations
onto the outgoing Stokes components of the transmitted light, is
given by
\begin{widetext}%
\begin{equation}
{\cal F}(p,s)=\frac{1}{\Delta(p,s)}%
\left(\begin{array}{cc}%
2\bar{\Xi}_3(\beta p\,s-\epsilon\kappa_2\Omega)\,+\,%
4\beta\epsilon\theta_z\bar{{\cal J}}_x\,\bar{\Xi}_3^2, &%
2\bar{\Xi}_3(\beta\Omega p\,+\,\epsilon\kappa_2\, s)\\%
2\bar{\Xi}_3(\epsilon\Omega p\,+\,\beta\kappa_2\, s), &%
-2\bar{\Xi}_3(\epsilon p\,s-\beta\kappa_2\Omega)\,-\,%
4\beta\epsilon\theta_y\bar{{\cal J}}_x\,\bar{\Xi}_3^2%
\end{array}\right)
\label{b.7}%
\end{equation}%
\end{widetext}%
The cross-transformation matrix ${\cal G}(p,s)$, which is
responsible for mapping the input Stokes components onto the
spatially distributed angular momentum fluctuations, is given by
\begin{widetext}%
\begin{equation}
{\cal G}(p,s)=\frac{1}{\Delta(p,s)}%
\left(\begin{array}{cc}%
-\bar{{\cal J}}_x(\theta_y\,p\,s-\theta_z\kappa_2\Omega)\,-\,%
2\epsilon\theta_y\theta_z\bar{{\cal J}}_x^2\bar{\Xi}_3, &%
\bar{{\cal J}}_x(\theta_y\kappa_2\,s+\theta_z\Omega\,p)\\%
\bar{{\cal J}}_x(\theta_z\kappa_2\,s+\theta_y\Omega\,p), &%
\bar{{\cal J}}_x(\theta_z\,p\,s-\theta_y\kappa_2\Omega)\,+\,%
2\beta\theta_y\theta_z\bar{{\cal J}}_x^2\bar{\Xi}_3
\end{array}\right)
\label{b.8}%
\end{equation}%
\end{widetext}%
Thus the Laplace images of the field and atomic variables become
fully defined.

To return into original space-time dependent representation it is
necessary to evaluate the integrals (\ref{6.10}). As we see from
(\ref{b.3})-(\ref{b.8}) the Laplace images of all the matrix
elements have polynomial structure and the return transform could
be found for any definite set of the external parameters, such as
$\kappa_2,\Omega,\ldots$ etc..

\end{document}